\documentclass[a4paper,11pt]{article}

\usepackage{iftex}
\ifPDFTeX\pdfoutput=1\fi
\usepackage{jheppub}
\usepackage{amsmath,amssymb,amsthm,mathtools,bm}
\usepackage{enumitem,array,mathrsfs,microtype,placeins}

\graphicspath{{figures/}}

\allowdisplaybreaks

\makeatletter
\def\@fpheader{\relax}
\makeatother

\renewcommand\baselinestretch{1.2}

\newcommand{\CFT}{\mathcal C}
\newcommand{\Hilb}{\mathcal H}
\newcommand{\Tr}{\operatorname{Tr}}
\newcommand{\Pf}{\operatorname{Pf}}
\newcommand{\id}{\mathbf 1}
\newcommand{\Z}{\mathcal Z}
\newcommand{\M}{\mathcal M}
\newcommand{\I}{\mathcal I}

\newcommand{\eps}{\epsilon}
\newcommand{\dd}{\mathrm d}
\newcommand{\ee}{\mathrm e}
\newcommand{\ket}[1]{|#1\rangle}

\newcommand{\Dket}[1]{|#1\rangle\!\rangle}

\newtheorem{proposition}{Proposition}[section]

\title{Stabilizer R\'enyi moments beyond entanglement}

\author[*,\dagger]{Vishnu Jejjala}
\author[\ddagger]{Joydeep Naskar}

\affiliation[*]{Mandelstam Institute for Theoretical Physics, School of Physics, University of the Witwatersrand, Johannesburg, South Africa}
\affiliation[\dagger]{National Institute for Theoretical and Computational Sciences (NITheCS), Gauteng, South Africa}
\affiliation[\ddagger]{Beijing Institute of Mathematical Sciences and Applications, Huaibei Town, Huairou District, Beijing 101408, China}

\emailAdd{v.jejjala@wits.ac.za}
\emailAdd{joydeepnaskar@bimsa.cn}

\abstract{
Stabilizer R\'enyi moments depend on a choice of operator basis and need not be determined by entanglement spectra.
In this paper, we study their spatial decay in critical quantum chains described by two-dimensional conformal field theories.
We derive an exact entanglement baseline for reductions of pure stabilizer states and a symmetry selection rule for the charged contribution at fixed support.
An exact Ising example shows that an on-site rotation changes the full decay exponent while preserving every subsystem entanglement spectrum.
For growing Ising intervals at index two, we obtain an exact Pfaffian representation and compare the charged amplitude with a candidate conformal boundary calculation.
The data support this prediction, but its identification with the lattice limit still requires endpoint matching and control of the order of limits.}

\keywords{Conformal and W Symmetry, Boundary Quantum Field Theory, Field Theories in Lower Dimensions, Discrete Symmetries}

\hypersetup{pdftitle={Stabilizer Renyi moments beyond entanglement in two dimensions},pdfauthor={Vishnu Jejjala, Joydeep Naskar}}

\begin{document}

\maketitle

\raggedbottom

\section{Introduction}
\label{sec:intro}

Stabilizer R\'enyi moments are constructed from Pauli or Weyl expectation values in a chosen operator basis.
For pure states, their logarithmic functionals are the stabilizer R\'enyi entropies, which are used to study quantum magic, or departures from the stabilizer formalism used in fault tolerant quantum computation~\cite{Veitch:2014rwa,Howard:2014qwa}.
They vanish on pure stabilizer states~\cite{Leone:2021sre} and are magic monotones for pure qubit states at integer moment order $\alpha\geq2$~\cite{LeoneBittel:2024}.
Related monotones and their operational properties were studied in Ref.~\cite{Haug:2023monotones}.

In critical systems, non-stabilizerness has been studied using mana in the Potts chain~\cite{White:2020zoz}, Bell measurements and Pauli sampling~\cite{Haug:2023bell,Haug:2023mps}, and measures of long range non-stabilizerness~\cite{Tarabunga:2023longrange,Korbany:2025longrange}.
The Bell basis representation gives a boundary conformal field theory (CFT) description of stabilizer moments.
In this description, boundary entropies and boundary condition changing (BCC) fields determine the scaling for a single region~\cite{Hoshino:2025sre,Hoshino:2025fusion}.
For free fermions, the mapping to participation entropies~\cite{Rajabpour:2025equiv} connects the problem to R\'enyi--Shannon boundary transitions~\cite{Stephan:2011rse,Stephan:2014mi,Alcaraz:2014mi}.
The decay for separated regions also depends on the exchanged fields and their couplings to the endpoints, which are not fixed by the single region data.
 
In this paper, we study stabilizer moments for separated regions of critical quantum chains described by two-dimensional CFTs.
We specify the local tensor product structure and the Pauli/Weyl basis, which together we call the \emph{stabilizer completion}.
Our observable is the joint moment divided by the product of the separate moments, so it equals one for a product state.
Since the reduced states are generally mixed even when the full chain is in a pure state, their stabilizer functionals are not pure state magic monotones.
We first need to determine what contribution can arise from entanglement alone.

An on-site unitary rotation preserves every subsystem entanglement spectrum, but need not preserve moments evaluated in a fixed operator basis.
We ask whether this difference changes only the amplitude or also the leading exponent of their spatial decay.
We also ask how the symmetry action on the operator basis determines this decay, and when the result at fixed lattice support survives a continuum limit.
This requires distinguishing two orders of limits.
One sends the separation to infinity at fixed lattice support.
The other grows both intervals at fixed geometric cross-ratio before taking the small cross-ratio limit.

We show that for a region $A$ whose reduced density matrix $\rho_A$ comes from a pure stabilizer state, the functional $\widehat\M_\alpha(A)$ defined in Section~\ref{sec:observable} obeys
\begin{equation}
 \widehat\M_\alpha(A) = \frac{S_2(A)}{\alpha-1} \,,\qquad
 S_2(A) = -\log\Tr\rho_A^2 \,,\qquad
 \alpha > 0 \,,\quad
 \alpha \neq 1 \,.
 \label{eq:intro-baseline}
\end{equation}
We use natural logarithms throughout.
For a pure global state, $S_2(A)$ measures entanglement with the complement, while for a mixed global state it only measures reduced state purity.
A nonzero value of $\widehat\M_\alpha$ in~\eqref{eq:intro-baseline} is therefore not evidence of magic.
Subtracting this entanglement baseline gives a nonnegative residual, but not a faithful mixed state magic measure.

For separated regions, we derive a selection rule when the on-site symmetry acts diagonally on the Weyl basis.
An operator is \emph{neutral} if conjugation leaves it invariant and \emph{charged} if it acquires a nontrivial phase.
In a symmetry invariant state, each charged one-region expectation vanishes, although oppositely charged operators can correlate across the two regions.
At fixed lattice support, the leading charged contribution to the normalized moment has power $\eta^{h_\text{ch}}$, where $\eta$ is the cross-ratio of the four interval endpoints and
\begin{equation}
 h_\text{ch} = 2\alpha\Delta^\text{ch}_{\min} \,,
 \label{eq:intro-charge}
\end{equation}
with $\Delta^\text{ch}_{\min}$ the smallest total scaling dimension with a nonzero coupling to the retained charged sector.
The result holds for real $\alpha>0$ under the stated pure power asymptotic assumptions.
Tests in the Ising CFT and the three-, four-, and five-state clock chains support this rule.
It does not, however, determine the leading power of the full ratio minus one when a neutral term dominates.

The dependence on the completion can be seen explicitly in the Ising model.
A particular on-site non-Clifford rotation preserves every subsystem entanglement spectrum but changes the cross-ratio exponent for two single sites from $\min\{\alpha/4,1\}$ to $1/4$ for integer $\alpha\geq2$.
This exact example shows that the operator basis can affect the leading exponent and not just its coefficient.
We also consider the Blume--Capel chain, where the symmetry permutes Weyl labels rather than multiplying each by a phase.

To study growing intervals, we use the infinite critical Ising chain.
We derive closed forms for the charged overlaps and express their sum as a quotient of finite ensembles.
At index two, this becomes a ratio of Pfaffians and permits evaluation through $128$ sites without enumerating Pauli strings.
An exact weighted pairing cancels the linear neutral energy response except for strings involving an endpoint.
The data at indices two and three support inverse length suppression of this remainder, although a uniform bound remains open.

At index two, we also calculate the charged amplitude for the lowest allowed junction of a candidate Ising orbifold boundary.
Its value is $1/4$, consistent with the growing interval lattice sequence.
Identifying it as the lattice limit still requires matching the normalized endpoint to this junction and controlling the order of limits.
The full continuum exponent also depends on the surviving neutral channels.

The organization of this paper is as follows.
In Section~\ref{sec:observable}, we define the observable and derive its entanglement baseline.
Section~\ref{sec:selection} gives the symmetry argument and the completion tests.
In Section~\ref{sec:boundary}, we discuss the Bell boundary and the assumptions needed for a continuum limit.
Section~\ref{sec:ising-exact} contains the exact Ising sums, the conditional CFT amplitude and the numerical tests.
We conclude with open questions and the relation to holography in Section~\ref{sec:discussion}.
Appendices~\ref{app:ising} and~\ref{app:boundary} contain the Ising identities and the boundary and junction calculations.
Appendix~\ref{app:notation} collects the notation.

\section{Stabilizer moments of subregions}
\label{sec:observable}

\subsection{Definitions and the entanglement baseline}

Consider a spin chain state $\rho$ with local dimension $d$ and a finite region $A$, with reduced density matrix $\rho_A=\Tr_{\bar A}\rho$ on a Hilbert space $\Hilb_A$ of dimension $D_A=d^{|A|}$.
We choose Weyl operators $W_{\bm u}$ satisfying
\begin{equation}
 \Tr(W_{\bm u}^{\dagger}W_{\bm v}) = D_A\delta_{\bm u,\bm v} \,,\qquad
 W_{\bm0} = \id_A \,,
 \label{eq:weyl}
\end{equation}
where $\id_A$ is the identity on $\Hilb_A$.
For the clock chains, the local Weyl operators are $w_{u_j}=X_j^{a_j}Z_j^{b_j}$, with $u_j=(a_j,b_j)\in\mathbb Z_d^2$ and $W_{\bm u}=\bigotimes_{j\in A}w_{u_j}$.
We use $ZX=\omega XZ$ with $\omega=\ee^{2\pi i/d}$.
Since all expressions below involve only magnitudes or phase covariant combinations, they also apply to composite $d$ without a half-phase convention~\cite{Gross:2006,Hostens:2005}.

The Weyl characteristic function, participation weights and moment are
\begin{equation}
 \chi_A(\bm u) = \Tr(\rho_AW_{\bm u}) \,,\qquad
 \Xi_A(\bm u) = D_A^{-1}|\chi_A(\bm u)|^2 \,,\qquad
 Z_\alpha(A) = \sum_{\bm u}\Xi_A(\bm u)^\alpha \,.
 \label{eq:moment}
\end{equation}
The moment is defined for every real $\alpha>0$.
Weyl orthogonality gives $\sum_{\bm u}\Xi_A(\bm u)=P_A$, where $P_A=\Tr\rho_A^2$ is the purity, so the normalized participation distribution is $p_A=\Xi_A/P_A$.
The corresponding logarithmic functional is
\begin{equation}
 \widehat\M_\alpha(A) = \frac{\log Z_\alpha(A)}{1-\alpha} - \log D_A \,,\qquad
 \alpha > 0 \,,\quad
 \alpha \neq 1 \,.
 \label{eq:mhat}
\end{equation}
For a pure state, this is the stabilizer R\'enyi entropy, while for a mixed state, substituting $\Xi_A=P_Ap_A$ gives
\begin{equation}
 \widehat\M_\alpha(A) = \frac{\alpha}{\alpha-1}S_2(A) + H_\alpha(p_A) - \log D_A \,,
 \label{eq:split}
\end{equation}
where $H_\alpha$ is the R\'enyi entropy of the probability distribution $p_A$.

\begin{proposition}[Entanglement baseline]
\label{prop:baseline}
If $\rho_A$ is the reduction of a pure stabilizer state, then
\begin{equation}
 \widehat\M_\alpha(A)=\frac{S_2(A)}{\alpha-1} \,.
 \label{eq:baseline}
\end{equation}
\end{proposition}
\begin{proof}
The stabilizer normal form is $\rho_A=D_A^{-1}\sum_{s\in S_A}s$, where $S_A$ is the subgroup supported in $A$~\cite{Fattal:2004stabilizer}.
Thus, $\Xi_A$ equals $1/D_A$ on $S_A$ and vanishes elsewhere.
Its total weight gives $|S_A|=D_AP_A$, so that $p_A$ is uniform on $D_AP_A$ labels.
Substituting $H_\alpha(p_A)=\log(D_AP_A)$ into~\eqref{eq:split} then gives the result.
\end{proof}

The same argument applies to normalized stabilizer code projectors, but not to arbitrary convex mixtures of stabilizer states.
For $0<\alpha<1$, the baseline is negative whenever $S_2>0$, and the nonnegativity statement below therefore concerns the \emph{residual},
\begin{equation}
 \mathcal A_\alpha(A) := \widehat\M_\alpha(A) - \frac{S_2(A)}{\alpha-1}
 = H_\alpha(p_A) - \log(D_AP_A) \,.
 \label{eq:residual}
\end{equation}
Since $|\chi_A|\leq1$, every probability is bounded by $1/(D_AP_A)$, and $H_\alpha(p_A)\geq H_\infty(p_A)\geq\log(D_AP_A)$.
It follows that $\mathcal A_\alpha\geq0$, with equality if and only if the participation distribution is flat.
In that case, the identity label has probability $1/(D_AP_A)$ and fixes the support size.

This residual is not a faithful mixed state magic measure.
For example, $\rho=\tfrac34|0\rangle\langle0|+\tfrac14|1\rangle\langle1|$ is a stabilizer mixture, but $\mathcal A_2=\log(20/17)>0$.
The residual measures departure from flat Weyl participation rather than from the stabilizer convex hull, and it is also distinct from the entanglement spectrum flatness studied in~\cite{Tirrito:2024flatness}.
At index two, it is the purity normalized $\widetilde M_2$ introduced in~\cite{Leone:2021sre} and used in CFT in~\cite{Hoshino:2025sre}.

\subsection{Separated regions and the two limits}

For two disjoint regions $A$ and $B$, it is convenient to work with the exact lattice ratio
\begin{align}
 F_{\alpha,\eps}(A,B) & := \frac{Z_\alpha(A\cup B)}{Z_\alpha(A)Z_\alpha(B)} \,,\nonumber\\
 \I_\alpha^\text{stab}(A:B) & := \widehat\M_\alpha(A) + \widehat\M_\alpha(B) - \widehat\M_\alpha(A\cup B)
 = \frac{\log F_{\alpha,\eps}(A,B)}{\alpha-1} \,.
 \label{eq:ratio}
\end{align}
Here, $\eps$ is the ultraviolet cutoff, equal to the lattice spacing for the uniform chains, and we write $I_2(A:B)=S_2(A)+S_2(B)-S_2(A\cup B)$ for the R\'enyi-$2$ mutual information.
For reductions of a single pure stabilizer state, Proposition~\ref{prop:baseline} gives $\I_\alpha^\text{stab}=I_2/(\alpha-1)$.
More generally, $Z_1$ is the purity, so $F_{1,\eps}=\ee^{I_2}$ and $(\alpha-1)\I_\alpha^\text{stab}\to I_2$, \textit{i.e.}, the functional has a pole at index one when $I_2\neq0$.

For intervals $A=[x_1,x_2]$ and $B=[x_3,x_4]$, with $x_1<x_2<x_3<x_4$ and $x_{ij}=x_i-x_j$, we define the cross-ratio
\begin{equation}
 \eta = \frac{x_{12}x_{34}}{x_{13}x_{24}} \,,\qquad
 0 < \eta < 1 \,.
 \label{eq:eta}
\end{equation}
On a circle of circumference $L$, distances are replaced by chord lengths $(L/\pi)\sin(\pi|x|/L)$.
For distant intervals of lengths $\ell_A,\ell_B$ and separation scale $r$, we have $\eta\simeq\ell_A\ell_B/r^2$.
We distinguish two orders of limits.
At fixed lattice support, we send $r\to\infty$.
The continuum limit instead increases $\ell_A/\eps$ and $\ell_B/\eps$ at fixed $\eta$ before taking the small-$\eta$ limit.
The charge sectors of~\eqref{eq:ratio} can be studied directly in the Weyl basis in the first limit.
Whether their exponents survive the second is a boundary CFT question.

\section{Symmetry selection and completion dependence}
\label{sec:selection}

\subsection{The charged exponent}

Consider an on-site symmetry $g=\prod_jX_j$ and a state satisfying $g\rho g^\dagger=\rho$.
A unique symmetry preserving ground state has this property, while in a degenerate ground state sector it must be imposed on the chosen state.
For $W_{\bm u}=\prod_jX_j^{a_j}Z_j^{b_j}$, we have
\begin{equation}
 g^\dagger W_{\bm u}g = \omega^{q(\bm u)}W_{\bm u} \,,\qquad
 q(\bm u) = \sum_jb_j\pmod d \,.
 \label{eq:charge}
\end{equation}
Since the partial trace preserves this on-site symmetry, cyclicity of the trace gives $\chi_A(\bm u)=\omega^{q(\bm u)}\chi_A(\bm u)$, so every charged expectation value vanishes.

We separate the joint moment into charge sectors,
\begin{equation}
 Z_\alpha^{(q_A,q_B)}(A,B) = \sum_{\substack{q(\bm u_A)=q_A\\ q(\bm u_B)=q_B}} \Xi_{A\cup B}(\bm u_A,\bm u_B)^\alpha \,.
 \label{eq:sectors}
\end{equation}
Only sectors with $q_A+q_B=0$ modulo $d$ survive.
The neutral sector $(0,0)$ has disconnected value $Z_\alpha(A)Z_\alpha(B)$, whereas each sector $(q,-q)$ with $q\neq0$ is purely connected, since both one region expectations vanish.

\begin{proposition}[Fixed support charged power]
\label{prop:charge}
Assume that the retained charged joint expectations have pure power large distance asymptotics with subleading remainders at fixed supports $A,B$.
Let $\Delta_q$ be the smallest scaling dimension with a nonzero joint leading amplitude for at least one retained label pair, including nonzero overlaps in both regions.
The normalized sector contribution then has leading power $\eta^{2\alpha\Delta_q}$ for real $\alpha>0$.
This pure power statement excludes leading multiplicative logarithms and zeros along separation subsequences.
For a sum of charged sectors,
\begin{equation}
 h_\text{ch}=2\alpha\Delta^\text{ch}_{\min} \,.
 \label{eq:hcharged}
\end{equation}
\end{proposition}
\begin{proof}
A charged joint label has no disconnected term.
Write $\chi_{A\cup B}=\Lambda_{AB}r^{-2\Delta_q}+o(r^{-2\Delta_q})$ for a pair with nonzero leading amplitude $\Lambda_{AB}$, including any degenerate fields in this amplitude.
Raising the magnitude to $2\alpha$ gives $r^{-4\alpha\Delta_q}$, or $\eta^{2\alpha\Delta_q}$ at fixed interval lengths, while the denominator is positive and independent of separation.
There are finitely many labels, so the summed remainder is still subleading.
The leading weights are nonnegative and cannot cancel.
\end{proof}

For the clock chain tests, we distinguish the all charged sum over $q=1,\ldots,d-1$ from the \emph{fundamental} sum $Z_\alpha^{\text{ch},1}$ over $\mathcal Q_1=\{1,-1\}\pmod d$.
The two sums agree for $d=2,3$ but differ for $d=4,5$, and at $d=2$ the set $\mathcal Q_1$ has only one element, which is counted once.

\subsection{The selection rule in the replica tensor}

The same selection rule has a replica interpretation.
For integer $\alpha\geq2$, we can write the moment as the operator insertion
\begin{equation}
 Z_\alpha(A) = \Tr\!\left[\rho^{\otimes2\alpha}\prod_{x\in A}Q_{\alpha,x}\right] \,,\qquad
 Q_{\alpha,x} = d^{-\alpha}\sum_{u_x}w_{u_x}^{\otimes\alpha}\otimes(w_{u_x}^\dagger)^{\otimes\alpha} \,.
 \label{eq:replica}
\end{equation}
Let $\bm v=(1,\ldots,1,-1,\ldots,-1)$, with $\alpha$ entries of each sign.
Conjugating the sheets by $\bigotimes_sX_s^{n_s}$ multiplies a charge-$q$ tensor term by $\omega^{q\bm v\cdot\bm n}$.
Since the Weyl tensor terms are linearly independent, the preserved subgroup and its annihilator are
\begin{equation}
 K_\alpha = \ker(\bm v\cdot)\simeq\mathbb Z_d^{2\alpha-1} \,,\qquad
 K_\alpha^\perp = \{q\bm v:q\in\mathbb Z_d\} \,.
 \label{eq:replica-symmetry}
\end{equation}
The map $\bm v\cdot$ is surjective, so this pairing also holds for composite $d$.
Suppose that the interval insertion, reference boundary and contracted endpoint multiplet preserve $K_\alpha$, so that each contracted interval insertion is a singlet.
A nonneutral exchanged state must then carry charge $q$ on every ket sheet and $-q$ on every bra sheet, and its lightest product has weight $2\alpha\Delta_q$, as in~\eqref{eq:hcharged}.
This identifies an allowed channel, but its continuum coefficient depends on the endpoint contraction.

For comparison, at $\alpha=1$ the charged contribution to the purity ratio has leading power $2\Delta^\text{ch}_{\min}$.
The logarithm of the full ratio is then the R\'enyi-$2$ mutual information, which does not separate into additive charge contributions.
More generally, the ordinary R\'enyi mutual information channel uses two sheets when its coefficient is nonzero, whereas the Bell channel in~\eqref{eq:replica-symmetry} uses all $2\alpha$ sheets.

\subsection{Symmetries that permute Weyl labels}
\label{sec:rotation}

What changes if the symmetry permutes Weyl labels instead?
We can test this in the Ising model without changing any entanglement spectrum.
Let us consider the infinite critical ground state $\ket\psi$ of $H=-\sum_j(X_jX_{j+1}+Z_j)$, and define
\begin{equation}
 R_y = \ee^{-i\pi Y/8} \,,\qquad
 \mathcal U=\bigotimes_j R_y \,,\qquad
 \ket{\psi'} = \mathcal U\ket\psi \,,\qquad
 H' = \mathcal U H\mathcal U^\dagger \,.
 \label{eq:rotation}
\end{equation}
The Hamiltonians are isospectral, and the reduced density matrices are related by a unitary on each region, so every subsystem entanglement spectrum is unchanged.
We keep the Pauli basis fixed when evaluating the two states.
The rotated symmetry is then the transversal Hadamard, whose local generator $R_yZR_y^\dagger=(X+Z)/\sqrt2$ exchanges $X$ and $Z$ and sends $Y$ to $-Y$.
It relates labels in orbits rather than assigning each label a phase.
By itself, it forces only odd labels fixed by the swap to vanish.
All odd-$Y$ expectations nevertheless vanish for this state, since both the ground state and $R_y$ are real.

For two single sites separated by $r$ lattice spacings, we set $\eta=r^{-2}$, $m_z=2/\pi$ and $e_0=m_z^2/2$.
Denoting the rotated state by primes, the one site unnormalized moment is $T'_\alpha=\sum_u|\chi'_A(u)|^{2\alpha}=1+2^{1-\alpha}m_z^{2\alpha}$, and the exact spin correlators give
\begin{equation}
 F'_{\alpha,\eps}-1 = \frac{\alpha(2\alpha-1)e_0^{2\alpha-2}\mathcal A_\sigma^2}{(T'_\alpha)^2} \eta^{1/4} + O(\eta^{1/2}) \,,\qquad
 \mathcal A_\sigma = 2^{1/12}\ee^{3\zeta'(-1)} \,,
 \label{eq:rotated-power}
\end{equation}
for every fixed integer $\alpha\geq2$, as derived in Appendix~\ref{app:rotation}.
Here, $\mathcal A_\sigma$ is the amplitude in $\langle X_0X_r\rangle\sim\mathcal A_\sigma r^{-1/4}$, and $\zeta$ is the Riemann zeta function.
In the unrotated state, the charged term has power $\alpha/4$ and the neutral energy term has power one, both with positive coefficients, and we therefore find
\begin{equation}
 h_\text{full}^\text{unrot} = \min\{\alpha/4,1\} \,,\qquad
 h_\text{full}^\text{rot} = 1/4 \,.
 \label{eq:completion-test}
\end{equation}
At $\alpha=2$, the powers differ by a factor of two despite identical entanglement spectra.
The rotated power emerges only at large separation because the higher powers have large relative coefficients.
For instance, at $\alpha=4$ a fit over $r=32,\ldots,256$ gives approximately $0.36$, and the local slope reaches $0.255$ only around $r=10^5$.

The spin-$1$ Blume--Capel chain gives another example of a symmetry that is not diagonal in the Weyl basis~\cite{Xavier:2011bc}.
We choose qutrit labels $j=0,1,2$ for $S^z=0,+1,-1$, with $X|j\rangle=|j+1\rangle$ and $Z|j\rangle=\omega^j|j\rangle$, so that the physical spin flip $C|j\rangle=|-j\rangle$ sends $X^aZ^b$ to $X^{-a}Z^{-b}$.
Writing $W_{\bm u}^{\pm}=(W_{\bm u}\pm W_{-\bm u})/2$, spin flip invariance removes mixed parity expectations, so a joint expectation is $E+O$, with $E=\langle W_{\bm u_A}^{+}W_{\bm u_B}^{+}\rangle$ and $O=\langle W_{\bm u_A}^{-}W_{\bm u_B}^{-}\rangle$.
The paired label $(-\bm u_A,\bm u_B)$ gives $E-O$, and the moment sum therefore cancels all terms linear in $O$.

For integer $\alpha\geq2$, suppose the odd part couples to the spin field and the even part has a nonzero disconnected background.
The first odd correction is then quadratic, with power $2\Delta_\sigma$, whereas an orbit with zero even background starts at $|O|^{2\alpha}$.
The neutral energy correction can still enter linearly.
Under these assumptions, the Ising candidate is $\min\{2\Delta_\sigma,\Delta_\varepsilon\}=1/4$.

\section{The Bell boundary and the continuum problem}
\label{sec:boundary}

\subsection{From the lattice insertion to boundary fields}

The fixed support exponent does not tell us whether its coefficient survives as the intervals grow.
To address this question, we rewrite the moment insertion as a boundary condition in the replicated theory.
We work at integer $\alpha\geq2$, and comparisons at real index are formal continuations unless stated otherwise.

Vectorizing an operator gives $\Dket X=\sum_{ij}X_{ij}|i\rangle|\bar j\rangle$, with inner product $\langle\!\langle X|Y\rangle\!\rangle=\Tr(X^\dagger Y)$.
The normalized Weyl vectors $\Dket{\widehat W_{\bm u}}=D_A^{-1/2}\Dket{W_{\bm u}}$ form a generalized Bell basis, in which $\Xi_A$ is the participation weight of $\Dket{\rho_A}$~\cite{Hoshino:2025sre,Hoshino:2025fusion}.
The local moment tensor is
\begin{equation}
 B_{\alpha,x} = \sum_{u_x}P_{u_x}^{\otimes\alpha} \,,\qquad
 P_u = \Dket{\widehat w_u}\langle\!\langle\widehat w_u| \,.
 \label{eq:bell-tensor}
\end{equation}
It imposes the same Bell label in every replica, and realigning its indices by $[\mathcal R(P)]_{il;jk}=P_{ij;kl}$ gives
\begin{equation}
 \mathcal R(P_u) = d^{-1}w_u\otimes w_u^\dagger \,,\qquad
 \mathcal R_\alpha(B_{\alpha,x}) = Q_{\alpha,x} \,,\qquad
 \mathcal R_\alpha(P_0^{\otimes\alpha}) = d^{-\alpha}\id^{\otimes2\alpha} \,,
 \label{eq:realignment}
\end{equation}
where $\mathcal R_\alpha$ applies $\mathcal R$ to each replica and permutes the factors into the order used in~\eqref{eq:replica}.
Thus, the identity Weyl projector on the complement becomes a transparent reference line with a local normalization.

For the vacuum of a uniform critical chain, we fold the $2\alpha$ ket and bra path integrals across the equal time line.
If the original CFT, denoted $\CFT$, is parity invariant with central charge $c$, the folded theory is $\CFT^{\otimes2\alpha}\otimes\overline{\CFT}^{\otimes2\alpha}$ with $c_\text{fold}=4\alpha c$.
We denote the measured Bell boundary on $A$ by $\Gamma_1$, the transparent reference on its complement by $\Gamma_0$, and the state preparation boundary by $\Gamma_2$.
Unfolding the reference strip gives
\begin{equation}
 \Hilb_{00} \simeq \Hilb_{\CFT}^{\otimes2\alpha} \,,\qquad
 Z_{00}(q) = \left[\sum_{i,j}M_{ij}\chi_i(q)\bar\chi_j(q)\right]^{2\alpha} \,,
 \label{eq:reference}
\end{equation}
where $\chi_i,\bar\chi_j$ are the chiral and antichiral characters, which should not be confused with the Weyl characteristic function $\chi_A$, and $M$ is the modular invariant multiplicity matrix~\cite{Recknagel:2002perm}.
A strip of width $L_\text{strip}$ unfolds to a circle of circumference $2L_\text{strip}$, so $q=\exp(-\pi\beta/L_\text{strip})$ for propagation length $\beta$.

Let $\gamma_A$ equal $\Gamma_1$ on $A$ and $\Gamma_0$ elsewhere, and define $\Z_\text{fold}^{(\alpha)}[\gamma;\Gamma_2]$ as the regulated Euclidean partition function on the preparation strip, with boundary condition $\gamma$ on the equal time edge and $\Gamma_2$ on the other edge.
The limit $\beta\to\infty$ selects the vacuum, and for a periodic chain the same notation denotes the cylinder path integral.
We can then write
\begin{equation}
 Z_\alpha(A) = \ee^{-S_\text{ct}^{(\alpha)}(A,\eps)} \frac{\Z_\text{fold}^{(\alpha)}[\gamma_A;\Gamma_2]}{\Z_\text{fold}^{(\alpha)}[\Gamma_0;\Gamma_2]} \,.
 \label{eq:partition}
\end{equation}
The term $S_\text{ct}^{(\alpha)}$ collects local line and endpoint counterterms.
Equation~\eqref{eq:partition} becomes a conformal boundary description if the measured and preparation boundaries have conformal renormalization group (RG) limits.
The Bell boundary need not be topological or a single copy Cardy state.
Appendix~\ref{app:zero-modes-lines} explains what additional annulus data a topological line interpretation would require.

For straight intervals, locality allows a counterterm proportional to their total length and a normalization at each endpoint~\cite{Billo:2016defects,Calabrese:2004eu}.
Homogeneity fixes the line density to a constant $\mu_\alpha/\eps$.
Each disjoint interval has one junction of each orientation, so the endpoints give a fixed constant.
Thus,
\begin{equation}
 -\log Z_\alpha(A) = \mu_\alpha\frac{\ell_A}{\eps} + c_\alpha^\partial N_{\partial A} + F_\alpha^\text{univ}(A;\eps) + o(1) \,.
 \label{eq:counterterms}
\end{equation}
Here, $\ell_A$ is the total interval length and $N_{\partial A}$ counts the endpoints.
The term $F_\alpha^\text{univ}$ contains the separation dependent logarithms and cross-ratio dependence, with an additive local convention that depends on the scheme.
The length and endpoint terms cancel in~\eqref{eq:ratio}, since a disjoint union adds both counts.
A conformal limit still requires control of the corrections from irrelevant defect and junction operators, and of possible logarithms from marginal operators~\cite{CardyCalabrese:2010corr}.

For example, distinct primary junction fields with weights $h_i>h_0$ have a vanishing cross term in the single interval two-point function, so their relative correction is of order $(\eps/\ell)^{2(h_i-h_0)}$.
A multi-junction correlator can instead have corrections of order $(\eps/\ell)^{h_i-h_0}$.
A defect perturbation of dimension $\widehat\Delta>1$ can contribute at first order, giving $(\eps/\ell)^{\widehat\Delta-1}$.
Bulk irrelevant fields give further powers, and descendants and endpoint shifts may add analytic terms.

An interval changes the boundary condition as $\Gamma_0\to\Gamma_1\to\Gamma_0$, so if the leading BCC field has weight $h_\alpha$, then
\begin{align}
 Z_\alpha(A) &= \mathcal N_\alpha\ee^{-\mu_\alpha\ell/\eps} (l_c/\eps)^{-2h_\alpha}[1+o(1)] \,,\nonumber\\
 \widehat\M_\alpha(A) &= m_\alpha\ell/\eps + \frac{2h_\alpha}{\alpha-1}\log(l_c/\eps) - \frac{\log\mathcal N_\alpha}{\alpha-1} + o(1) \,,
 \label{eq:interval}
\end{align}
where $l_c$ is the chord length and $m_\alpha=\mu_\alpha/(\alpha-1)-\log d$.
The endpoint normalization $\mathcal N_\alpha$ is not an Affleck--Ludwig boundary entropy~\cite{Cardy:1989ir,Affleck:1991tk}.
Writing $g_a$ for the vacuum overlap of $\Gamma_a$, the latter instead enters the full system constant as $-\log(g_1/g_0)/(\alpha-1)$, after cancelling the common preparation boundary overlap.

\subsection{Two intervals and the neutral channel}
\label{sec:neutral}

For two intervals, let $\psi_{ab}$ denote a BCC field changing $\Gamma_a$ to $\Gamma_b$.
If the lattice ratio converges at fixed $\eta$ to the scalar contraction of the leading endpoint fields, its limit is
\begin{equation}
 \mathfrak F_\alpha(\eta) = \frac{\langle\psi_{01}(x_1)\psi_{10}(x_2)\psi_{01}(x_3)\psi_{10}(x_4)\rangle}{\langle\psi_{01}(x_1)\psi_{10}(x_2)\rangle \langle\psi_{01}(x_3)\psi_{10}(x_4)\rangle} \,.
 \label{eq:four-point}
\end{equation}
Fusing the endpoints of each interval exchanges fields in $\Hilb_{00}$ and gives the block expansion
\begin{equation}
 \mathfrak F_\alpha(\eta) = \sum_{p,a}A_{p,a}\mathcal V_p^{(c_\text{fold})}(\eta)
 = 1 + A_\star\eta^{h_\star} + o(\eta^{h_\star}) \,,
 \label{eq:blocks}
\end{equation}
where $a$ labels multiplicities and $\mathcal V_p$ is a diagonal Virasoro block with external weights $h_\alpha$.
It includes the descendants of the exchanged primary of weight $h_p$ and is normalized so that $\mathcal V_0(0)=1$ and $\mathcal V_p(\eta)\sim\eta^{h_p}$ for a nonvacuum primary.
For $A_\star\neq0$, this gives $\I_\alpha^\text{stab}\sim A_\star\eta^{h_\star}/(\alpha-1)$.
The exchanged weight $h_\star$ belongs to the reference channel, whereas $h_\alpha$ is the endpoint weight.

If $\psi_{10}=\psi_{01}^\dagger$ in a unitary reflection positive boundary theory, the complete module coefficients are sums of squared operator product expansion (OPE) amplitudes and are therefore nonnegative.
For $h_\alpha\neq0$, the vacuum module then contains $2h_\alpha^2\eta^2/c_\text{fold}$, which other positive norm modules cannot cancel.
However, the realignment of the lattice tensor does not establish this adjoint scalar contraction.
If the endpoints form a non-adjoint multiplet, we must specify its matrix contraction in place of~\eqref{eq:four-point}, since assigning signs to scalar module coefficients is not enough.

Fusing the inner endpoints instead exchanges $\Hilb_{11}$ and gives the crossing relation
\begin{equation}
 \sum_{p,a}A_{p,a}\mathcal V_p(\eta) = \left(\frac{\eta}{1-\eta}\right)^{2h_\alpha}
 \sum_{q,b}B_{q,b}\mathcal V_q(1-\eta) \,.
 \label{eq:crossing}
\end{equation}
The identity in this channel gives $\I_\alpha^\text{stab}=2h_\alpha\log[1/(1-\eta)]/(\alpha-1)+O(1)$ for one approaching junction pair, and a complementary bipartition of a circle has two such pairs, hence twice this coefficient.

The neutral sector explains why the full small-$\eta$ exponent need not equal the charged exponent.
For a neutral label, we write $\chi_{A\cup B}=z+\delta\chi(r)$, where $z=\chi_A\chi_B$ is the disconnected value.
Since the first variation of $|z+\delta\chi|^{2\alpha}$ is $2\alpha|z|^{2\alpha-2}\operatorname{Re}(z^*\delta\chi)$, an energy field contribution $\delta\chi\sim r^{-2\Delta_\varepsilon}$ can give a term of power $\eta^{\Delta_\varepsilon}$, independent of $\alpha$.

We normalize the energy field by $\langle\varepsilon(x)\varepsilon(0)\rangle=|x|^{-2\Delta_\varepsilon}$.
Its coefficient $C_{\varepsilon,A}(\bm u)$ is defined through the long distance expansion $W_{\bm u}=\chi_A(\bm u)\id+C_{\varepsilon,A}(\bm u)\varepsilon(x_A)+\cdots$ about the interval center $x_A$.
The normalized linear energy coefficient is then
\begin{equation}
 \mathcal K_\varepsilon(A) = \sum_{\bm u}|\chi_A(\bm u)|^{2\alpha-2}\chi_A(\bm u)^*C_{\varepsilon,A}(\bm u) \,,\qquad
 \nu_A = \frac{\mathcal K_\varepsilon(A)}{T_\alpha(A)\ell_A^{\Delta_\varepsilon}} \,,\qquad
 A_1 = 2\alpha\operatorname{Re}(\nu_A\nu_B) \,,
 \label{eq:neutral-overlap}
\end{equation}
where $T_\alpha(A)=\sum_{\bm u}|\chi_A(\bm u)|^{2\alpha}$.
The subscript in $A_1$ refers to the linear neutral response rather than to the moment index, and its $\alpha$ dependence is suppressed.
Pairing opposite Weyl labels shows that $\mathcal K_\varepsilon$ is real, since the phase in $W_{-\bm u}=\ee^{i\theta(\bm u)}W_{\bm u}^\dagger$ cancels between the two factors in each summand.

For infinite critical Ising, Appendix~\ref{app:neutral-pairing} uses the Cauchy covariance to pair neutral labels with equal moment weights and opposite energy overlaps, cancelling all terms not involving a chosen endpoint.
The number and weights of the remaining endpoint strings grow with the interval, so their sum still needs a bound; Section~\ref{sec:charged-amplitude} examines its numerical behavior.

The algebraic pairing does not identify the full Kramers--Wannier (KW) action on the Bell junction.
Duality on interior operators leaves endpoint terms~\cite{Hoshino:2025fusion}, and its continuum action also depends on the junction and zero mode projection~\cite{Chang:2019tdl,Seiberg:2024kw}.
Appendix~\ref{app:zero-modes-lines} illustrates how zero modes can restore a neutral power absent from the oscillator correlator.

If the linear term vanishes, the quadratic energy response has power $2\Delta_\varepsilon$.
Its Taylor coefficient is nonnegative, but higher terms in the expansion of $\chi$ may contribute at the same power, and these terms must be combined into complete conformal modules before applying the positivity argument above.
In the Ising model, for instance, the two energy and vacuum stress contributions both have power two.

\subsection{Conditions for a continuum exponent}
\label{sec:conditions}

To obtain a continuum exponent from Proposition~\ref{prop:charge}, we need to control the remainder as the intervals grow.
Let $\lambda=(\ell_A/\eps,\ell_B/\eps)$ and denote a normalized charged sum by $F_\alpha^\text{ch}(\eta;\lambda)$.
A sufficient condition is
\begin{equation}
 \begin{aligned}
 F_\alpha^\text{ch}(\eta;\lambda) &= \mathcal C_\alpha(\lambda)\eta^{h_\text{ch}}+R_\alpha(\eta;\lambda) \,,\\
 \mathcal C_\alpha(\lambda) & \longrightarrow\mathcal C_\alpha\in(0,\infty) \,,
 & |R_\alpha(\eta;\lambda)| & \leq K\eta^{h_\text{ch}+\delta} \,,
 \end{aligned}
 \label{eq:uniform}
\end{equation}
for constants $K,\delta>0$ and a range $0<\eta<\eta_0$ independent of $\lambda$, together with the existence of the fixed-$\eta$ limit.
This uniform control allows the fixed support power to survive the growing interval limit, whereas positivity at each fixed support does not supply the bound.
These conditions are sufficient, not necessary.

For the full exponent, we will use the following assumptions:
\begin{enumerate}[label=(A\arabic*),leftmargin=3.2em,itemsep=1pt,topsep=3pt]
\item The measured boundary and its endpoint contraction have a conformal limit, with their scalar or multiplet character specified.
\item The charged sector satisfies~\eqref{eq:uniform}.
\item The single sheet neutral coefficient tends to zero, while the remaining neutral expansion has a fixed-$\eta$ limit and a uniform higher power remainder.
\item No neutral power below $h_\text{cand}$ in~\eqref{eq:candidate} survives, and the aggregate coefficient of $\eta^{h_\text{cand}}$ is nonzero after combining all contributions at that power.
\end{enumerate}
In (A3), all other lower powers must be included before imposing the remainder bound.
Under (A1)--(A4), the charged field, the two sheet energy product and the vacuum module then give the candidate
\begin{equation}
 h_\star = h_\text{cand} := \min\{2\alpha\Delta^\text{ch}_{\min},\,2\Delta_\varepsilon,\,2\} \,.
 \label{eq:candidate}
\end{equation}
The stress power $2$ is present only if it survives the contraction in (A1).
An adjoint scalar contraction with $h_\alpha\neq0$ guarantees a positive coefficient, and otherwise this must be checked under (A4).

For example, the parafermion CFTs of the Fateev--Zamolodchikov $\mathbb Z_d$ chains have
\begin{equation}
 c = \frac{2(d-1)}{d+2} \,,\qquad
 \Delta_{\sigma_1}=\frac{d-1}{d(d+2)} \,,\qquad
 \Delta_\varepsilon=\frac4{d+2} \,.
 \label{eq:parafermions}
\end{equation}
The order field is the lightest charged primary, while the first thermal field is odd under duality and the next even thermal field is heavier than the two sheet energy product~\cite{Fateev:1985para,Mong:2014}.
Comparing these weights for $d=2,3,4,5$ gives
\begin{equation}
 h_\text{cand}^{(d)} = \min\!\left\{\frac{2\alpha(d-1)}{d(d+2)},\frac8{d+2}\right\} \,,\qquad
 \alpha_\times = \frac{4d}{d-1} \,.
 \label{eq:clock-candidate}
\end{equation}
The equality points are $8,6,16/3,5$, respectively, and they describe crossovers only if (A1)--(A4) hold in the relevant boundary phase.
A charged weight reaching one in the reference channel does not imply an instability of the measured boundary; that would require an allowed perturbation in $\Hilb_{11}$.
For the non-diagonal Potts invariant, the measured replica boundary and its annulus multiplicities remain to be identified~\cite{Affleck:1998potts,Mong:2014}.

At finite support, the single sheet neutral coefficient may remain nonzero.
For equal intervals of length $\ell$, we then write $F_{\alpha,\eps}-1\simeq\mathcal C_\alpha(\ell)\eta^{h_\text{ch}}+A_1(\ell)\eta^{\Delta_\varepsilon}$, using $\mathcal C_\alpha(\ell)$ for the charged coefficient in~\eqref{eq:uniform}.
When $h_\text{ch}>\Delta_\varepsilon$, the two terms have equal magnitude at $\eta_\times=|A_1/\mathcal C_\alpha|^{1/(h_\text{ch}-\Delta_\varepsilon)}$, and fits on opposite sides of this scale can give different effective powers.
If the powers coincide, the coefficients simply add, so the crossover scale is undefined while the moment remains regular.

\subsection{Ising boundary data}
\label{sec:ising-boundary}

Ref.~\cite{Hoshino:2025sre} constructs a replica symmetric candidate for the measured Ising boundary.
In this case, the transparent annulus~\eqref{eq:reference} is $[\chi_0^2+\chi_{1/2}^2+\chi_{1/16}^2]^{2\alpha}$, which contains the all sheet spin state of weight $\alpha/4$.
At index two, the reference--measurement gluing introduces one Dirichlet--Neumann oscillator, whose vacuum weight is $1/16$.
If this vacuum survives the orbifold projection and is the leading junction selected by the lattice endpoint, it predicts
\begin{equation}
 h_2 = \frac1{16} \,,\qquad
 c_2^{\log} = 2h_2 = \frac18 \,.
 \label{eq:ising-bcc}
\end{equation}
In Appendix~\ref{app:orbifold-junction}, we check this projection for the elementary boundary pair with zero displacement and zero Wilson line, and find that within this family the resolution compatible with projector positivity retains one lowest junction.
If the lattice endpoint selects it, the index two logarithm agrees with the pure stabilizer baseline, since the logarithmic coefficient of $S_2$ is $c/4=1/8$~\cite{Calabrese:2004eu,Calabrese:2009qy}.
This agreement concerns only the logarithm and does not identify the full reduced state or its magic.
At higher index, the reference--measurement junction requires a separate analysis, which the fully Dirichlet comparison in Appendix~\ref{app:boundary} does not supply.

With $g_0=1$, the replica symmetric construction and the full system free fermion mapping also give the boundary overlap~\cite{Hoshino:2025sre,Rajabpour:2025equiv},
\begin{equation}
 g_1 = \begin{cases} \sqrt\alpha \,, & 2 \leq \alpha \leq 4 \,,\\ 2 \,, & \alpha > 4\end{cases} \,,\qquad
 \widehat\M_\alpha(\mathrm{full}) = m_\alpha L - \frac{\log g_1}{\alpha-1} + o(1) \,.
 \label{eq:ising-g}
\end{equation}
Here, $L$ is the number of sites, and the two branches meet at the R\'enyi--Shannon boundary transition $\alpha_c=4$~\cite{Stephan:2011rse,Stephan:2014mi}.
A full system overlap does not determine the local junction coefficients on either branch, and the marginal point may have logarithmic corrections.
We explain the orbifold normalization in Appendix~\ref{app:boundary}.
Section~\ref{sec:kink} considers a kink hypothesis for the locked branch alongside the lattice coefficients.

\section{Exact Ising coefficients and numerical tests}
\label{sec:ising-exact}

In infinite critical Ising, we can compute the charged coefficient and the linear neutral response without first identifying the boundary.
We use these exact lattice expressions to test the continuum assumptions and compare the index-two coefficient with the candidate CFT boundary.
Throughout this section and in Appendix~\ref{app:ising}, lengths and separations are in lattice units, \textit{i.e.}, $\ell$ counts sites in an interval and $L$ counts sites on a ring.

\subsection{The charged coefficient as a finite ensemble quotient}
\label{sec:pfaffian}

The Ising Majorana covariance is a Cauchy matrix, whose determinant identities turn sums over Pauli strings into sums over subsets of doubled lattice nodes.
At index two, these sums admit a Pfaffian evaluation.
Let $T_\alpha(\ell)$ be the unnormalized moment of an $\ell$-site interval defined in Section~\ref{sec:neutral}, and let $C_{\bm u}$ be the limit of $\langle W_{\bm u}X_R\rangle/\langle X_{\ell-1}X_R\rangle$ as $R\to\infty$ for a charged string.
We then write
\begin{equation}
 \Sigma_\alpha(\ell) = \sum_{\bm u\ \text{charged}}|C_{\bm u}|^{2\alpha} \,,\qquad
 U_\alpha(\ell) = \frac{\Sigma_\alpha(\ell)}{T_\alpha(\ell)} \,.
 \label{eq:U}
\end{equation}
For equal intervals, the coefficient of the fixed support charged power is
\begin{equation}
 \mathcal C_\alpha(\ell) := \lim_{\eta\to0}\eta^{-\alpha/4} \frac{Z_\alpha^{\text{ch},1}(A,B)}{Z_\alpha(A)Z_\alpha(B)}
 = \mathcal A_\sigma^{2\alpha}\ell^{-\alpha/2}U_\alpha(\ell)^2 \,.
 \label{eq:charged-coefficient}
\end{equation}
This is also the leading coefficient of the full ratio, but only when the charged channel leads.

To rewrite the sums, let $\mathsf X_\ell=\{0,1,\ldots,\ell-1\}$, $\mathsf Y_\ell=\mathsf X_\ell+\tfrac12$, $\Lambda_\ell=\mathsf X_\ell\cup \mathsf Y_\ell$ and $V(S)=\prod_{u<v\in S}(v-u)$, and introduce the finite ensemble
\begin{equation}
 \mathscr Z_N^{(\alpha)}[w] = \sum_{\substack{S\subset\Lambda_\ell\\ |S|=N}} V(S)^{2\alpha}\prod_{u\in S}w(u)^{2\alpha} \,,
 \label{eq:ensemble}
\end{equation}
with
\begin{equation}
 w_0(u) = \frac1{\Gamma(u+\tfrac12)\Gamma(\ell-u+\tfrac12)} \,,\qquad
 w_1(u) = \frac1{\Gamma(u+\tfrac12)\Gamma(\ell-u)} \,,\qquad
 \mathcal G_\ell = \prod_{n=1}^{\ell}\frac{\Gamma(n-\tfrac12)}{\Gamma(n)} \,,
 \label{eq:weights}
\end{equation}
where $\Gamma$ denotes the Euler gamma function, and the Cauchy complement identities in Appendix~\ref{app:cauchy} give
\begin{equation}
 U_\alpha(\ell) = \left(\frac{\mathcal G_\ell}{\sqrt\pi}\right)^{2\alpha} \frac{\mathscr Z_{\ell-1}^{(\alpha)}[w_1]}{\mathscr Z_\ell^{(\alpha)}[w_0]} \,.
 \label{eq:ensemble-ratio}
\end{equation}
This is the open line counterpart of the balanced doubled root bijection in~\cite{Khasseh:2026fugacity}.
The charged ensemble has $\ell-1$ particles rather than $\ell$, with a changed one body weight $w_1/w_0=\Gamma(\ell-u+\tfrac12)/\Gamma(\ell-u)$.
Both changes are needed to represent the endpoint contribution.
Identifying the continuum insertion of this quotient therefore requires more than the full system participation mapping~\cite{Rajabpour:2025equiv}.

At index two, the fourth power of the Vandermonde allows a Pfaffian evaluation.
We set $p_i(u)=u^i$ and define the $2N\times2N$ skew matrix
\begin{equation}
 \mathsf M^{(N)}_{ij}[w] = \sum_{u\in\Lambda_\ell}w(u)^4 \bigl[p_i(u)p'_j(u)-p'_i(u)p_j(u)\bigr] \,,\qquad
 0 \leq i,j < 2N \,.
 \label{eq:skew-moments}
\end{equation}
The confluent Vandermonde identity gives $\mathscr Z_N^{(2)}[w]=\Pf\mathsf M^{(N)}[w]$ with no extra factorial, and consequently
\begin{equation}
 U_2(\ell) = \left(\frac{\mathcal G_\ell}{\sqrt\pi}\right)^4 \frac{\Pf\mathsf M^{(\ell-1)}[w_1]}{\Pf\mathsf M^{(\ell)}[w_0]} \,,\qquad
 \mathcal C_2(\ell) = \frac{\mathcal A_\sigma^4}{\ell}U_2(\ell)^2 \,.
 \label{eq:pfaffian-ratio}
\end{equation}
Here, the empty Pfaffian is one.
This is the standard confluent construction, whose weighted circular version appears in~\cite{Khasseh:2026fugacity}.
Appendix~\ref{app:pfaffian} gives the proof and derives the linear neutral response by differentiating a tilted Pfaffian.
Evaluating the matrices takes $O(\ell^3)$ arithmetic operations, but their ill-conditioning requires increasing precision.

\begin{table}[tb]
\centering\small\setlength{\tabcolsep}{6pt}
\begin{tabular}{lrrrr}
\hline
Quantity & $\ell=4$ & $\ell=8$ & $\ell=12$ & $\ell=16$\\
\hline
$\mathcal C_2$ & $0.17991$ & $0.19816$ & $0.20677$ & $0.21207$\\
$\mathcal C_3$ & $0.07677$ & $0.08365$ & $0.08763$ & $0.09038$\\
$\mathcal C_4$ & $0.02357$ & $0.02245$ & $0.02205$ & $0.02184$\\
$10^3\mathcal C_5$ & $5.772$ & $4.160$ & $3.464$ & $3.049$\\
$10^3\mathcal C_6$ & $1.272$ & $0.6524$ & $0.4428$ & $0.3363$\\
$10^5\mathcal C_8$ & $5.627$ & $1.422$ & $0.6358$ & $0.3589$\\
$10^6\mathcal C_{10}$ & $2.433$ & $0.3051$ & $0.09055$ & $0.03824$\\
\hline
\end{tabular}
\caption{\textsf{ Exact charged coefficients for equal Ising intervals, using the length convention in~\eqref{eq:charged-coefficient}.
The coefficients at $\alpha=2,3$ increase slowly, while those at $\alpha=4$ decrease mildly.
At higher index, the decrease at these sizes is largely accounted for by the identity and single string terms.}}
\label{tab:coefficients}
\end{table}

Table~\ref{tab:coefficients} gives the exact coefficients through $\ell=16$.
At $\alpha=2,3,4$, the sequences are compatible with nonzero limits, although extrapolations depend on the fit ansatz.
The ratio $U_\alpha$ avoids the additional choice of assigning a continuum length to an $\ell$-site block.
We first examine index two, where the Pfaffian permits larger intervals, and return to the higher-index coefficients in Section~\ref{sec:kink}.

\subsection{A conditional CFT amplitude at index two}
\label{sec:charged-amplitude}

For comparison with CFT, we express the exact coefficient as the interval's response to a distant order field probe.
At index two, the local lattice tensor is itself a positive projector,
\begin{equation}
 Q_{2,x} = \frac14\bigl(\id^{\otimes4}+X_x^{\otimes4} + Y_x^{\otimes4}+Z_x^{\otimes4}\bigr)
 = \frac14(\id^{\otimes4}+X_x^{\otimes4})(\id^{\otimes4}+Z_x^{\otimes4}) \,.
 \label{eq:Q2-projector}
\end{equation}
We write $\mathscr D_A=\prod_{x\in A}Q_{2,x}$ and let $\mathcal P(A)=\{\id,X,Y,Z\}^{\otimes\ell}$ denote the Pauli strings on $A$, with overall phases omitted.
In the four copy vacuum,
\begin{equation}
 \frac{\langle\mathscr D_A\prod_{a=1}^4X_r^{(a)}\rangle}{\langle\mathscr D_A\rangle} = \frac{\sum_{P\in\mathcal P(A)}\langle P X_r\rangle^4}{T_2(\ell)} \,.
 \label{eq:charged-probe-v2}
\end{equation}
Since $X_r$ is odd under the Ising spin flip symmetry, only charged strings contribute to the numerator.
Large separation selects their leading spin field coupling.
We normalize the Ising spin field by $\langle\sigma(z)\sigma(0)\rangle=|z|^{-1/4}$, so that $X_x\sim\mathcal A_\sigma^{1/2}\sigma(x)$ in lattice units and the exchanged field $\mathcal S=\sigma_1\sigma_2\sigma_3\sigma_4$ has unit two-point normalization and dimension $1/2$.
Its dimensionless far field coupling, which fixes the normalization of the amplitude without assuming a four-endpoint correlator, is
\begin{equation}
 b_\ell = \frac{\mathcal A_\sigma^2U_2(\ell)}{\sqrt\ell} \,,\qquad
 \mathcal C_2(\ell) = b_\ell^2 \,.
 \label{eq:charged-coupling}
\end{equation}

To evaluate this coupling for the candidate boundary, we pair the four Ising copies into two orbifold bosons with action $\sum_i\int(\partial\phi_i)^2/(2\pi)$ and period $2\pi$, for which the normalized bosonization is
\begin{equation}
 \mathcal S=2\cos\phi_1\cos\phi_2 = \cos(\sqrt2\phi_+) + \cos(\sqrt2\phi_-) \,,\qquad
 \phi_\pm = \frac{\phi_1\pm\phi_2}{\sqrt2} \,.
 \label{eq:charged-bosonization}
\end{equation}
Each cosine on the right has dimension $1/2$ and two-point coefficient $1/2$.
In the candidate gluing of~\cite{Hoshino:2025sre}, the relative field $\phi_-$ is Dirichlet at zero on the interval while $\phi_+$ remains transparent, and both fields are transparent outside the interval.
This is the reference--measurement junction of Section~\ref{sec:ising-boundary}.
Our transparent reference differs from the fully Dirichlet reference used in that work.

Only the relative field cosine then has a nonzero one-point function, and mapping the plane slit along $[0,\ell]$ to the exterior disc gives
\begin{equation}
 \frac{\langle\mathscr D_{[0,\ell]}\mathcal S(r)\rangle_\text{CFT}}{\langle\mathscr D_{[0,\ell]}\rangle_\text{CFT}} = \left[\frac{\ell}{4r(r-\ell)}\right]^{1/2}
 = \frac{\sqrt\ell}{2r}\,[1+O(\ell/r)] \,,\qquad
 r > \ell \,.
 \label{eq:slit-response}
\end{equation}
We derive this expression in Appendix~\ref{app:orbifold-junction}, where we also show that the orbifold projection leaves the charged coupling of the retained lowest junction equal to $1/2$.
Subject to the matching conditions discussed below, the boundary calculation therefore predicts
\begin{equation}
 \lim_{\ell\to\infty}\mathcal C_2(\ell) = \frac14 \,,\qquad
 U_2(\ell) \sim \frac{\sqrt\ell}{2\mathcal A_\sigma^2} \,.
 \label{eq:conditional-amplitude}
\end{equation}
To infer~\eqref{eq:conditional-amplitude} from this junction calculation, we still need to establish that the projector reaches this boundary, that the normalized endpoint insertion converges to its lowest junction, and that the far field projection converges uniformly as $\ell$ grows.
These are the parts of (A1)--(A2) needed here.
The second condition does not follow from a positive overlap at every finite cutoff, since such overlaps can still vanish in the limit.
The third is needed because convergence only at fixed $r/\ell$ would not justify the far field first order of limits in~\eqref{eq:charged-coefficient}.
The value $1/4$ here is an amplitude multiplying $\eta^{1/2}$, not the rotated Ising exponent in~\eqref{eq:completion-test}.

\begin{figure}[tb]
\centering\includegraphics[width=\textwidth]{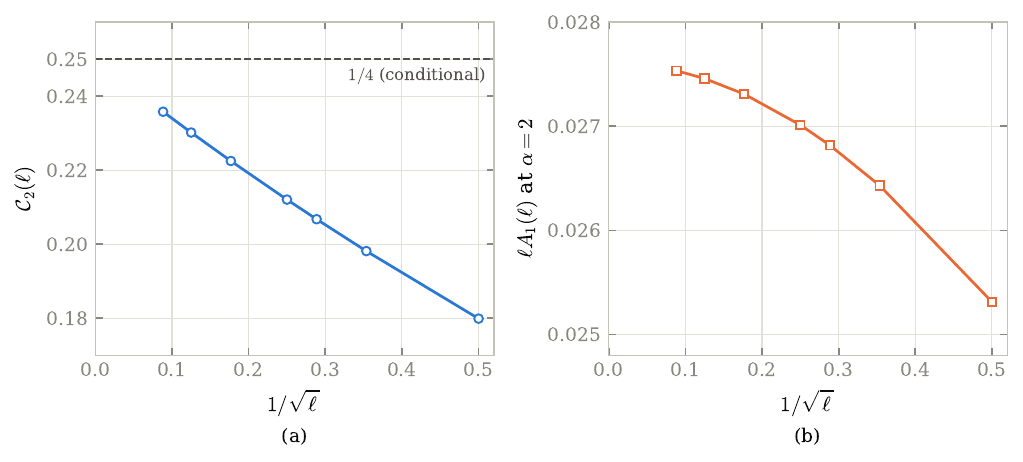}
\caption{\textsf{ Growing equal intervals in infinite critical Ising at index two.
(a) The charged coefficient~\eqref{eq:pfaffian-ratio}, with the conditional CFT amplitude shown as a dashed line.
(b) The rescaled linear neutral coefficient~\eqref{eq:neutral-pfaffian}, whose near constancy supports $A_1\propto1/\ell$.
Solid segments join adjacent sizes.}}
\label{fig:growing-ising}
\end{figure}

Figure~\ref{fig:growing-ising} compares this prediction with the exact Pfaffian coefficients at $\ell=4,8,12,16,32,64,128$ and also shows the linear neutral coefficient.
The charged sequence reaches $\mathcal C_2(128)=0.2357979$, about $6\%$ below the conditional value $1/4$.
Fits to $c_0+c_1/\sqrt\ell+c_2/\ell$ over the largest three, four and five sizes give $c_0=0.24995,0.24991,0.24988$.
Their spread is not an extrapolation error, since the correction ansatz itself is unproved.

For the linear neutral coefficient we find $\ell A_1(128)=0.0275335$, compared with $0.0270122$ at $\ell=16$, which extends the evidence for $1/\ell$ suppression of the endpoint remainder.
At $\alpha=3$, direct sums through $\ell=16$ give $\ell A_1\simeq0.00773$ with the same apparent scaling.
Neither sequence, however, determines the complete power two neutral coefficient.

The Pfaffian evaluations agree with the direct charged sums through $\ell=16$ to within $2\times10^{-14}$.
We stabilize the moment matrices by centering and rescaling the nodes.
Each size is computed at $80+2\ell$ decimal digits, and repeating the calculation with $40$ additional digits leaves all $50$ retained digits unchanged.

\subsection{Higher indices and the kink hypothesis}
\label{sec:kink}

Above the boundary transition $\alpha_c=4$ in~\eqref{eq:ising-g}, we can ask whether the charged insertion is a kink of the locked boundary.
In the associated height theory, the boundary locking field and its dual unit kink have folded weights
\begin{equation}
 x_\text{lock} = 4/\alpha \,,\qquad
 x_\text{kink} = \alpha/4 \,.
 \label{eq:kink-weight}
\end{equation}
To obtain the kink weight, we impose a step between adjacent minima on each side of the probability slit and add the classical actions.
For action $\alpha\kappa\int(\partial\varphi)^2/(4\pi)$ and period $2\pi r_0$, a locking harmonic $p$ gives $\alpha_c=p^2/(2\kappa r_0^2)$, and a step $2\pi r_0/p$ then has $x_\text{kink}=\alpha/\alpha_c$.

To apply this result to the subsystem moment, we need a further assumption:
\begin{enumerate}[label=(A\arabic*),leftmargin=3.2em,itemsep=1pt,topsep=3pt]
\setcounter{enumi}{4}
\item The normalized charged insertion couples to the unit kink with finite nonzero amplitude, and its endpoint factors introduce no term of equal or lower power.
\end{enumerate}
Integrating one kink over each distant interval then gives $\ell_A\ell_B r^{-2x_\text{kink}}$.
Including the cutoff dependence and expressing lengths in lattice units, the charged coefficient for equal intervals would obey
\begin{equation}
 \mathcal C_\alpha(\ell) \propto \ell^{2-\alpha/2} \,,\qquad
 \alpha > 4 \,.
 \label{eq:kink-prediction}
\end{equation}
Thus, (A5) would violate the positive limit condition in (A2) on the locked branch.
The kink is marginal at $\alpha=4$, and below four the dilute kink expansion about the locked boundary does not apply.
Even if~\eqref{eq:kink-prediction} holds, removing this power from the continuum expansion requires uniform control of the remainder as in~\eqref{eq:uniform}.
It would not establish that the entire charged sector vanishes at fixed cross-ratio.

For the coefficients in Table~\ref{tab:coefficients} at $\alpha=5,6,8,10$, the quantity $\ell^{\alpha/2-2}\mathcal C_\alpha$ changes by less than $6\%$ between $\ell=4$ and $16$.
This is compatible with the power in~\eqref{eq:kink-prediction}, but identity dominance gives the same behavior.
The identity contributes one to $T_\alpha$, while the $\ell$ single-$X_j$ overlaps each have unit magnitude.
If these terms dominate, $U_\alpha\simeq\ell$ and $\mathcal C_\alpha\sim\ell^{2-\alpha/2}$, and indeed $U_\alpha(16)/16\simeq(0.99,1.02,1.01,1.00)$ at those four indices.
For a twelve-site block, the sum $T_\alpha(\ell)=D_A^\alpha Z_\alpha(A)$ similarly falls from about $60$ at $\alpha=2$ to $1.4$ at $\alpha=5$.
This sum is not an effective number of strings.

However, identity dominance cannot persist to arbitrarily large support at fixed finite $\alpha$, since the diagonal Cauchy minors give
\begin{equation}
 T_\alpha(\ell) \geq (1+m_z^{2\alpha})^\ell \,,\qquad
 m_z = 2/\pi \,,
 \label{eq:lower-bound}
\end{equation}
as proved in Appendix~\ref{app:cauchy}.
This bound reaches $\ee$ at a size of about $92$ for $\alpha=5$ and $8364$ for $\alpha=10$.
These sizes are not computed crossover lengths, but they show why much larger blocks may be needed.

\subsection{Charge sectors in Ising and clock chains}
\label{sec:tests}

We use Ising and the interacting $\mathbb Z_d$ chains, $d=3,4,5$, to test different aspects of the selection rule.
Infinite chain Ising correlations avoid ring effects, while the clock chains probe different charged dimensions.
The latter are taken at the Fateev--Zamolodchikov point~\cite{Fateev:1985para},
\begin{equation}
 H = -\sum_j\sum_{n=1}^{d-1}\frac{(Z_j^\dagger Z_{j+1})^n+X_j^n}{\sin(\pi n/d)} \,,
 \label{eq:FZ}
\end{equation}
with periodic boundary conditions, where the $d=3$ model is the Potts chain.
Our Ising convention $H=-\sum_j(X_jX_{j+1}+Z_j)$ is related to the $d=2$ clock convention by a local Hadamard transformation, which leaves the moments invariant.

\begin{figure}[tb]
\centering\includegraphics[width=\textwidth]{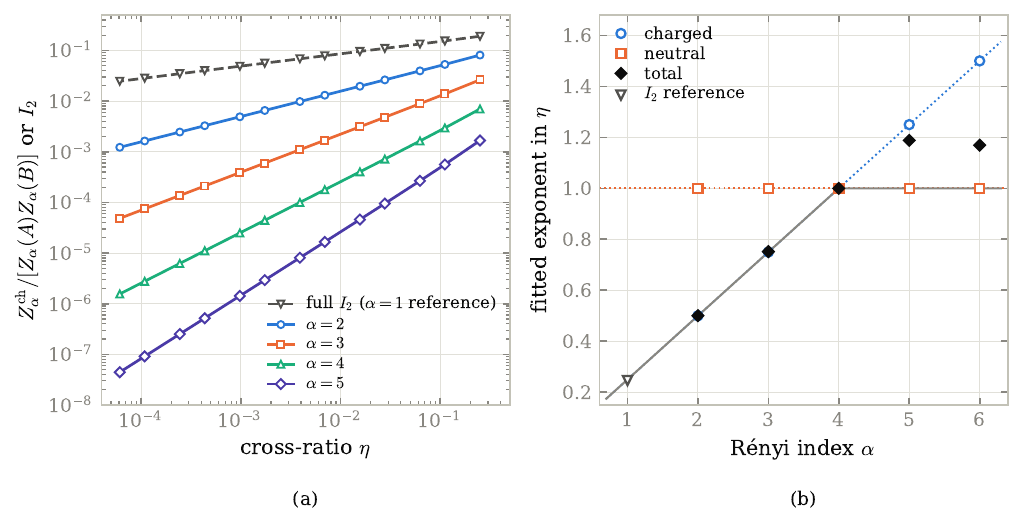}
\caption{\textsf{ Separated regions in infinite critical Ising.
(a) The charged ratio for two length-two intervals follows $\eta^{\alpha/4}$.
(b) For one site intervals, the charged and single sheet neutral powers are $\alpha/4$ and $1$, the grey curve is their fixed support minimum, and above $\alpha=4$ the full ratio fits mix these lattice terms.
In both panels, the $\alpha=1$ entries show the full R\'enyi-$2$ mutual information, rather than a charged component or the singular functional $\widehat\M_\alpha$ at index one.}}
\label{fig:sectors}
\end{figure}

Figure~\ref{fig:sectors} separates the Ising contributions.
The charged power follows analytically from the order field correlator, so its numerical fit mainly checks the implementation.
For one site regions, the neutral term has power one.

\begin{figure}[tb]
\centering\includegraphics[width=0.75\textwidth]{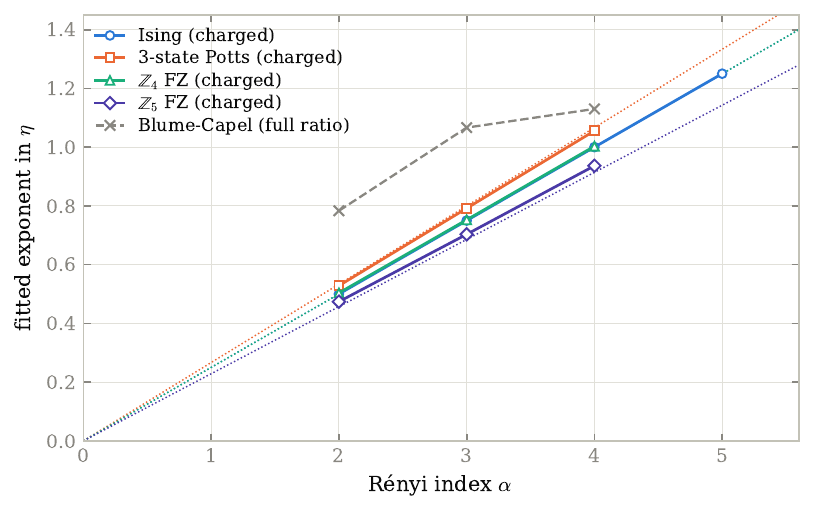}
\caption{\textsf{ Fixed support exponents for the clock models and the Blume--Capel control.
The dotted guides use the exact charged CFT dimensions in~\eqref{eq:hcharged}, and the grey control points are mean full ratio powers over separation windows.
The Ising charged points are fits to the six smallest cross-ratios, $\eta\leq1/576$, while the interacting clock points use their full separation windows.
Table~\ref{tab:clock} reports the full window charged slopes, including Ising, together with their window or size sensitivities.}}
\label{fig:exponents}
\end{figure}

For the clock chains, we use individual sites separated by $r=2,\ldots,\lfloor L/2\rfloor$.
Fitting the logarithm of the fundamental charged ratio against $\log\eta$ gives an exponent at each index.
We then fit $h_\text{ch}(\alpha)=h_\text{int}+s_\text{ch}\alpha$, using $\alpha=2,3,4,5$ for Ising and $\alpha=2,3,4$ for the other models.
Figure~\ref{fig:exponents} and Table~\ref{tab:clock} compare the slope $s_\text{ch}$ with $2\Delta_\text{CFT}$ and a reference value $2\Delta_\text{ref}$.
For Ising, the reference is the exact infinite chain value, whereas for the interacting clock chains it is fitted independently from the order field correlator on the same ring.
The same ring ratio can cancel a common finite size drift, while the exact CFT dimension provides an independent benchmark.
Extended fit and size sequences are given in the accompanying code documentation.

\begin{table}[tb]
\centering\small\setlength{\tabcolsep}{4.5pt}
\begin{tabular}{llcccc}
\hline
Model & Support & $s_\text{ch}$ & $2\Delta_\text{CFT}$ & $2\Delta_\text{ref}$ & $s_\text{ch}/(2\Delta_\text{ref})$\\
\hline
Ising & $\ell=2$, $L=\infty$ & $.2526(17)$ & $.2500$ & $.2500$ & $1.010(7)$\\
Potts & one site, $L=12$ & $.2637(15)$ & $.2667$ & $.2639(14)$ & $.9992(4)$\\
$\mathbb Z_4$ & one site, $L=10$ & $.2497(8)$ & $.2500$ & $.2508(5)$ & $.9958(14)$\\
$\mathbb Z_5$ & one site, $L=8$ & $.2313(9)^\dagger$ & $.2286$ & $.2341(3)^\dagger$ & $.9880(27)^\dagger$\\
\hline
\end{tabular}
\caption{\textsf{ Charged sector slopes and scaling dimensions.
The reference is the exact value for infinite chain Ising and the same ring order field fit for the other models.
Parentheses give the maximum shift on removing the one or two shortest separations while retaining at least three points, while daggered entries instead give the $L=8\to10$ change about the displayed $L=8$ value.
These are window or size sensitivities rather than statistical or controlled thermodynamic extrapolation errors.}}
\label{tab:clock}
\end{table}

The charged slopes agree with the prediction at the percent level, and the size sequences, which extend to $L=14,12,10$ for $d=3,4,5$, respectively, have same ring ratios approaching unity.
The free intercepts are small compared with the measured exponents, although the Potts and $\mathbb Z_4$ window spreads do not resolve them as zero.
The neutral contribution remains appreciable: at $\alpha=2$ it amounts to $12$--$21\%$, $24$--$32\%$, and $38$--$45\%$ of the all charged contribution for $d=3,4,5$ over the fitted separations.
These single site tests therefore support the charged law.

For the non-Weyl control, we use $H_\text{BC}=-\sum_jS_j^zS_{j+1}^z-\lambda\sum_jS_j^x+D\sum_j(S_j^z)^2$ at $D=0$, $\lambda=1.326$, and $L=12$, near the Ising critical point~\cite{Xavier:2011bc}.
If we incorrectly imposed the qutrit phase charge rule, it would forbid sectors whose measured weight reaches $0.0597$, whereas the correct charge conjugation orbit sum cancels the linear odd contribution to numerical precision.
Averaging full ratio powers over contiguous windows of at least three separations within $r=2,\ldots,6$ gives approximately $0.783,1.067,1.130$ at $\alpha=2,3,4$ (Figure~\ref{fig:exponents}).
These are full-ratio fits, not tests of the charged-sector law~\eqref{eq:hcharged}, and the ring is too small to test the asymptotic candidate $1/4$.

\subsection{The single interval check}

\begin{figure}[tb]
\centering\includegraphics[width=.67\textwidth]{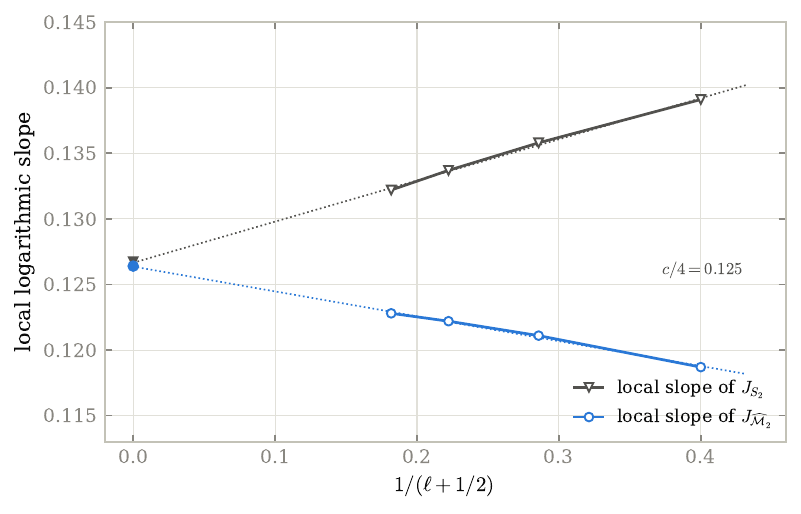}
\caption{\textsf{ Local logarithmic slopes of $J_M(\ell)$ in infinite critical Ising for $M=S_2$ (triangles) and $M=\widehat\M_2$ (circles), plotted at $1/(\ell+1/2)$ for $\ell=2,3,4,5$.
The dotted lines are separate linear fits, and the filled symbols at the origin are their extrapolations rather than computed sizes, giving $0.1267$ for the R\'enyi-$2$ entanglement and $0.1264$ for the stabilizer series, to be compared with $c/4=1/8$.
The two finite size drifts have opposite signs.}}
\label{fig:baseline}
\end{figure}

The candidate Ising junction in~\eqref{eq:ising-bcc} predicts that the single interval logarithm at index two equals the entanglement baseline.
To test this, we define $J_M(\ell)=2M(\ell)-M(2\ell)$ for adjacent equal intervals, taking $M$ to be either $\widehat\M_2$ or $S_2$.
Its local slope is the change from $\ell$ to $\ell+1$ divided by $\log[(\ell+1)/\ell]$.
Figure~\ref{fig:baseline} shows both sequences.

Linear fits to the slope difference give intercepts of about $-3.0\times10^{-4}$ using all four points and $-6.1\times10^{-5}$ using the last three.
Thus no additional logarithm is resolved at $\alpha=2$, although the sequences are too short to bound its continuum coefficient at sub-percent accuracy.
Finally, full system fits on $L=6,8,10,12$ agree with the constants in~\eqref{eq:ising-g} within $0.7$--$2.9\%$ for $\alpha=2,\ldots,6$, and we include the marginal index four only as a finite size comparison.

\FloatBarrier
\section{Discussion}
\label{sec:discussion}

Stabilizer moments distinguish spatial scaling that subsystem entanglement spectra do not determine.
The rotated Ising calculation establishes this exactly.
An on-site rotation changes the full decay exponent, not merely its amplitude.
In a symmetry invariant state, the selection rule~\eqref{eq:hcharged} fixes the charged power at fixed support when the symmetry is diagonal in the Weyl basis, although neutral terms can determine the full decay exponent.
The baseline~\eqref{eq:baseline} identifies the entanglement contribution present even for reductions of pure stabilizer states, but subtracting it does not produce a faithful mixed state magic measure.

For growing Ising intervals, the exact charged sums and weighted neutral pairing lead to Pfaffian expressions for the charged coefficient and the remaining linear neutral response at index two.
Their evaluation through $128$ sites supports the candidate charged amplitude $1/4$ and inverse-length suppression of the neutral remainder.
Neither lattice limit is proved.
The charged amplitude requires endpoint matching and uniform control of the far field limit, and the neutral remainder requires a bound on the endpoint-containing strings.
The prediction for the full continuum exponent additionally requires (A3)--(A4).
At higher Ising indices, the available coefficients do not distinguish the alternative kink hypothesis from finite-support identity dominance.
For the higher clock completions, both the measured boundary and a weighted neutral pairing remain unidentified.

Long range magic provides a further test of the fixed cross-ratio limit.
In our conventions, the quantity studied in~\cite{Tarabunga:2023longrange} is
\begin{equation}
 L_\text{LR} = \mathcal A_2(A\cup B)-\mathcal A_2(A)-\mathcal A_2(B)
 = I_2(A:B)-\I_2^\text{stab}(A:B) \,.
 \label{eq:longrange}
\end{equation}
That work reported logarithmic growth for two quarter system intervals in periodic critical Ising, with chord cross-ratio $1/2$, whereas~\eqref{eq:four-point} assumes a finite limit at fixed cross-ratio.
Our exact sequence at $\eta=1/4$ gives $L_\text{LR}=0.0819,0.1088,0.1223,0.1307$ for $\ell=1,2,3,4$.
These sizes do not distinguish saturation from logarithmic growth, and a direct comparison requires growing intervals in the same geometry as~\cite{Tarabunga:2023longrange}.

The charge argument also extends to multipartite moments.
The three region inclusion--exclusion combination of $\widehat\M_\alpha$ cancels local line and endpoint terms and involves a normalized six-point function with three cross-ratios in a conformal description~\cite{Malvimat:2026syk}.
A label charged on all three regions requires nonzero $q_A,q_B,q_C$ with $q_A+q_B+q_C=0\pmod d$.
There is no such triple for $d=2$, while $(1,1,d-2)$ works for every $d\geq3$.
This is not a monogamy theorem for qubits, since the logarithmic connected quantity can contain products of charged pair sectors.

It remains to ask whether mana, robustness, or a measure optimized over basis choices retains the spatial scaling information of our fixed-basis residual $\mathcal A_\alpha$.

\subsection{Relation to holography}

What would a holographic description of these moments require?
The pure stabilizer baseline involves $S_2$ at every moment index $\alpha$.
For the ordinary R\'enyi entropy $S_n(A)=\log\Tr\rho_A^n/(1-n)$, we define the refined entropy $\widetilde S_n=n^2\partial_n[(n-1)S_n/n]$~\cite{Dong:2016fnf}, and integrating gives $S_2=2\int_1^2\dd n\,\widetilde S_n/n^2$.
The semiclassical contribution of the baseline is therefore a weighted integral of cosmic brane areas, rather than the area of one brane with tension set by $\alpha$.
Moreover, the usual area prescriptions and their quantum corrections~\cite{Ryu:2006bv,Hubeny:2007xt,Faulkner:2013ana,Engelhardt:2014gca} do not determine the completion dependent residual.

A bulk calculation must also retain the all sheet replica structure of the charged channel in~\eqref{eq:replica-symmetry} and, as the rotated Ising example shows, information about the encoding.
This requirement is compatible with quantum error correction in bulk reconstruction~\cite{Harlow:2016vwg}, obstructions to state dependent area operators in exact stabilizer codes~\cite{Cao:2023fcm}, and proposals relating magic to the coupling between matter and geometry~\cite{Cao:2024gdi,Cao:2026codes}.
Obstructions also arise for exact stabilizer descriptions of local QFT vacua~\cite{Benedetti:2026}.\footnote{
Ref.~\cite{Shaposhnik:2026inf} excludes type III central summands in the local algebras of its canonical infinite lattice stabilizer states, which obstructs the QFT algebraic structure in that setting without establishing a no-go theorem for general continuum scaling limits.
For its Araki--Woods construction, type III is equivalent to unbounded min-relative entropy of magic optimized over arbitrary unitaries on each side of a fixed bipartition, so basis dependence within that bipartition is removed, unlike for our fixed frame moments.}

We have not identified bulk objects for the three Bell boundaries, and our calculation gives neither an area equality nor an area bound for $\mathcal A_\alpha$.
Entropy cone inequalities~\cite{Hayden:2011ag} also cannot be applied to this fixed basis functional as they stand, since its $\alpha\to1$ behavior is not the von Neumann entropy.
Before seeking a bulk prescription, one needs the endpoint contraction and the fixed cross-ratio coefficients, particularly at the marginal Ising index four and on the locked branch, and we leave these questions for future work.

\paragraph{Note added:}
While this work was being completed, Ref.~\cite{Chu:2026euw} introduced an average of Pauli moments over independent single qubit rephasings, taken before the logarithm.
This average removes local phase reference dependence but not dependence on arbitrary local basis rotations, whereas we keep the Pauli/Weyl frame fixed.
Rephasings that commute with the symmetry preserve charge sectors, so to retain~\eqref{eq:hcharged} after averaging, one must also check that the averaged leading coefficient is nonzero and control the fixed support asymptotics under the phase integral.
The rotation in Section~\ref{sec:rotation} is not a rephasing and therefore lies outside this average.

\paragraph{Code availability:}
Our numerical codes and implementation details can be found in this \href{https://github.com/naskarjoydeep/stabilizer-moments}{\texttt{Github} repository}.

\acknowledgments
The authors would like to thank Charles Cao for email correspondence.
V.J.\ thanks the NSF Institute for Artificial Intelligence and Fundamental Interactions (IAIFI) and the Department of Physics at Northeastern University for hospitality during his sabbatical during which much of this research was undertaken.
J.N.\ thanks the APS DGRAV and FIP for travel awards to attend the Global Physics Summit 2026 where part of this work was done.
Generative AI models from Anthropic and OpenAI were used to generate and debug numerical scripts.
The authors are solely responsible for the conclusions.
V.J.\ is supported by the South African Research Chairs Initiative of the Department of Science, Technology, and Innovation and the National Research Foundation (grant 78554).
J.N.\ is supported by the Beijing Institute of Mathematical Sciences and Applications.

\appendix
\section{Exact Ising identities}
\label{app:ising}

\subsection{Cauchy overlaps and the complement construction}
\label{app:cauchy}

The Jordan--Wigner transformation and Wick's theorem express the nonzero one-block Pauli expectations as minors of $G_{ij}=-[\pi(j-i+1/2)]^{-1}$.
For ordered sets $I,J\subset\{0,\ldots,\ell-1\}$ with $|I|=|J|=k$, we define $D_{IJ}=\det G_{I,J}$, and the Cauchy determinant identity gives
\begin{equation}
 |D_{IJ}| = \pi^{-k}\frac{V(I)V(J)}{P(I,J)} \,,\qquad
 P(I,J) = \prod_{i\in I,j\in J}|j-i+\tfrac12| \,.
 \label{eq:cauchy-minor}
\end{equation}
Here, empty products equal one, and summing $|D_{IJ}|^{2\alpha}$, including the empty minor, gives $T_\alpha$.

For a charged overlap, we apply Wick's theorem to $W_{\bm u}X_R$ and divide by $\langle X_{\ell-1}X_R\rangle$ before taking $R\to\infty$.
The remote factors then cancel, leaving
\begin{equation}
 |C_{IJ}| = \frac{\pi^{-k}}{\sqrt\pi}\frac{V(I)V(J)}{P(I,J)}\prod_{i\in I}\frac{\Gamma(\ell-i+\tfrac12)}{\Gamma(\ell-i)}\prod_{j\in J}\frac{\Gamma(\ell-j-\tfrac12)}{\Gamma(\ell-j)} \,,\qquad
 |I| = k \,,\quad
 |J| = k+1 \,,
 \label{eq:charged-overlap}
\end{equation}
where $I,J$ are the Majorana sets of $W_{\bm u}X_R$ restricted to the block.
The Jordan--Wigner tail complements the string's own sets, for which the condition is instead $|I_{\bm u}|=|J_{\bm u}|+1$, so restricting the covariance matrix to the disconnected sites without this tail misses the charged sector.
At $\ell=2$, for example, the four nonzero leading overlaps are $1,1,1/2,1/2$.

To use the doubled nodes of Section~\ref{sec:ising-exact}, we define $\mathsf X_I=\{i:i\in I\}$, $\mathsf Y_J=\{j+\tfrac12:j\in J\}$ and $J^c=\{0,\ldots,\ell-1\}\setminus J$.
The complement map $S(I,J)=\mathsf X_I\cup \mathsf Y_{J^c}$ maps neutral minors bijectively to $\ell$-particle subsets and charged overlaps to $(\ell-1)$-particle subsets~\cite{Khasseh:2026fugacity}.
Factorizing the Vandermonde of $\mathsf Y_\ell$ gives
\begin{align}
 |D_{IJ}| &= V(\mathsf Y_\ell)V(S)\prod_{u\in S}w_0(u) \,,\label{eq:neutral-term}\\
 |C_{IJ}| &= \frac{V(\mathsf Y_\ell)\mathcal G_\ell}{\sqrt\pi} V(S)\prod_{u\in S}w_1(u) \,.\label{eq:charged-term}
\end{align}
The factors follow from $\prod_{j=0}^{\ell-1}|i-j-\tfrac12|=\Gamma(i+\tfrac12)\Gamma(\ell-i+\tfrac12)/\pi$ and $\prod_{j'\neq j}|j-j'|=j!(\ell-1-j)!$.
Taking the $2\alpha$th powers and summing over subsets gives
\begin{equation}
 T_\alpha = V(\mathsf Y_\ell)^{2\alpha}\mathscr Z_\ell^{(\alpha)}[w_0] \,,\qquad
 \Sigma_\alpha=\left[\frac{V(\mathsf Y_\ell)\mathcal G_\ell}{\sqrt\pi}\right]^{2\alpha} \mathscr Z_{\ell-1}^{(\alpha)}[w_1] \,,
 \label{eq:ensemble-identities}
\end{equation}
and hence~\eqref{eq:ensemble-ratio}.
Since the distant order field limit defining $C_{\bm u}$ has already been taken, these identities do not give the full moment at finite cross-ratio.

For the denominator bound, we set $I=J$ in~\eqref{eq:cauchy-minor}, which gives
\begin{equation}
 |D_{II}| = m_z^{|I|}\prod_{i<j\in I}\frac{(j-i)^2}{(j-i)^2-\tfrac14}
 \geq m_z^{|I|} \,.
 \label{eq:diagonal-bound}
\end{equation}
Keeping these terms in the positive sum and summing over all subsets $I$ gives $(1+m_z^{2\alpha})^\ell$, as in~\eqref{eq:lower-bound}.

\subsection{The weighted neutral cancellation}
\label{app:neutral-pairing}

For a nonempty neutral minor, we define $v_{IJ}=\boldsymbol1^TG_{I,J}^{-1}\boldsymbol1$, where $\boldsymbol1$ is the vector of $k$ ones.
It satisfies the inverse sum identity
\begin{equation}
 v_{IJ} = -\pi\left(\sum_{j\in J}j-\sum_{i\in I}i+\frac{k}{2}\right) \,,\qquad
 |I| = |J| = k \,,\qquad
 v_{\varnothing,\varnothing} = 0 \,.
 \label{eq:energy-inverse-sum}
\end{equation}
To derive this identity, solve $\mathsf C\mathsf b=\boldsymbol1$ for $\mathsf C_{ij}=1/(y_j-x_i)$.
The rational function $f(z)=\sum_j\mathsf b_j/(y_j-z)$ then obeys $1-f(z)=\prod_i(z-x_i)/\prod_j(z-y_j)$.
Comparing the coefficients of $z^{-1}$ at infinity gives $\sum_j\mathsf b_j=\sum_jy_j-\sum_ix_i$.
Substituting $x_i=i$, $y_j=j+1/2$ and $G=-\mathsf C/\pi$ gives~\eqref{eq:energy-inverse-sum}.

For blocks with origins separated by $r$, let $(I_A,J_A)$ and $(I_B,J_B)$ be their local minor indices, and set $I_{AB}=I_A\cup(r+I_B)$, $J_{AB}=J_A\cup(r+J_B)$.
Since the leading cross-covariances are $\mp\boldsymbol1\boldsymbol1^T/(\pi r)$, with the vector lengths fixed by the two minors, the Schur complement gives
\begin{equation}
 \frac{D_{I_{AB}J_{AB}}}{D_{I_AJ_A}D_{I_BJ_B}} = 1 + \frac{v_{I_AJ_A}v_{I_BJ_B}}{\pi^2r^2} + o(r^{-2}) \,.
\end{equation}
The linear energy coefficient for equal intervals is consequently
\begin{equation}
 A_1(\ell) = \frac{2\alpha}{\pi^2\ell^2} \left(\frac{\sum_{I,J}|D_{IJ}|^{2\alpha}v_{IJ}}
 {T_\alpha(\ell)}\right)^2 \,.
 \label{eq:neutral-exact}
\end{equation}
The sum runs over equal cardinality sets.
It describes the linear response of neutral labels with nonzero disconnected expectation, not the entire neutral moment.

For pairs with $0\notin I$, consider
\begin{equation}
 I' = \{\ell-i:i\in I\} \,,\qquad
 J' = \{\ell-1-j:j\in J\} \,.
 \label{eq:neutral-involution}
\end{equation}
This involution reverses each Cauchy denominator and preserves the Vandermonde magnitudes, so that $|D_{I'J'}|=|D_{IJ}|$.
Equation~\eqref{eq:energy-inverse-sum} gives $v_{I'J'}=-v_{IJ}$.
The complete weighted terms therefore cancel for every real $\alpha>0$, and a fixed point has $v_{IJ}=0$.
If $0\in I$, on the other hand, the reflected set contains the excluded site $\ell$.
Hence,
\begin{equation}
 \sum_{I,J}|D_{IJ}|^{2\alpha}v_{IJ} = \sum_{\substack{I,J\\0\in I}}|D_{IJ}|^{2\alpha}v_{IJ} \,.
 \label{eq:neutral-endpoint}
\end{equation}

\subsection{The Pfaffian and its neutral derivative}
\label{app:pfaffian}

For $S=\{u_1<\cdots<u_N\}$, the confluent Vandermonde identity gives
\begin{equation}
 \det[p(u_1),p'(u_1),\ldots,p(u_N),p'(u_N)] = V(S)^4 \,,\qquad
 p(u) = (1,u,\ldots,u^{2N-1})^T \,.
 \label{eq:confluent-vandermonde}
\end{equation}
This identity follows by splitting each node in the ordinary Vandermonde into a nearby pair, dividing by the intra-pair differences and taking their coincident limits.
The top coefficient of the $N$th exterior power of $\sum_u w(u)^4p(u)\wedge p'(u)$, divided by $N!$, is $\Pf\mathsf M^{(N)}[w]$.
Since repeated nodes vanish and the $N!$ orderings of each subset cancel that denominator,~\eqref{eq:confluent-vandermonde} then gives $\mathscr Z_N^{(2)}[w]$, which proves~\eqref{eq:pfaffian-ratio}.

The neutral response follows from the same matrix.
Under $S=\mathsf X_I\cup \mathsf Y_{J^c}$,
\begin{equation}
 \mathfrak d_\ell(S) := \sum_{u\in S}u-\frac{\ell^2}{2}
 = \sum I-\sum J-\frac{k}{2} = \frac{v_{IJ}}{\pi} \,.
 \label{eq:neutral-displacement}
\end{equation}
Multiplying the neutral ensemble summand by $\exp[t\mathfrak d_\ell(S)]$ gives the Pfaffian matrix $\mathsf M(t)$ of~\eqref{eq:skew-moments} at $N=\ell$, with $w_0(u)^4$ replaced by $w_0(u)^4\exp[t(u-\ell/2)]$.
Differentiating $(\Pf\mathsf M)^2=\det\mathsf M$ then gives
\begin{equation}
 \overline{\mathfrak d}_\ell = \frac12\Tr[\mathsf M(0)^{-1}\mathsf M'(0)] \,,\qquad
 A_1(\ell)=\frac4{\ell^2}\overline{\mathfrak d}_\ell^{\,2} \,,\quad (\alpha=2) \,.
 \label{eq:neutral-pfaffian}
\end{equation}
The bar denotes the neutral ensemble average.
Thus both the charged coefficient and the linear neutral response can be evaluated from skew moments and their first derivative.

\subsection{The rotated two site moment}
\label{app:rotation}

We write $c_{\mu\mu}(r)=\langle\sigma_0^\mu\sigma_r^\mu\rangle$ for $\mu=x,y,z$, where $(\sigma^x,\sigma^y,\sigma^z)=(X,Y,Z)$ are the lattice Pauli operators.
Expectations in the rotated state use $R_y^\dagger XR_y=(X+Z)/\sqrt2$ and $R_y^\dagger ZR_y=(Z-X)/\sqrt2$.
This is the opposite conjugation to $R_yZR_y^\dagger$, which transforms the symmetry generator.
In the unrotated state, the on-site Ising symmetry sets mixed $XZ$ correlators to zero, and reality removes single $Y$ correlators.
With $E=c_{zz}/2$ and $O=c_{xx}/2$, the rotated $XX,ZZ$ expectations are $E+O$, whereas $XZ,ZX$ are $E-O$, and the sixteen-label sum becomes
\begin{equation}
 F'_{\alpha,\eps}-1 = \frac{2|E+O|^{2\alpha}+2|E-O|^{2\alpha}-4e_0^{2\alpha}+|c_{yy}|^{2\alpha}}{(T'_\alpha)^2} \,.
 \label{eq:rotated-exact}
\end{equation}
The correlators are
\begin{align}
 c_{xx}(r) &= \left(\frac2\pi\right)^r \prod_{j=1}^{r-1}\left(1-\frac1{4j^2}\right)^{j-r}
 \sim\mathcal A_\sigma r^{-1/4} \,,\\
 c_{zz}(r) &= m_z^2+\frac4{\pi^2(4r^2-1)} \,,\qquad
 c_{yy}(r) = -\frac{c_{xx}(r)}{4r^2-1} \,.
 \label{eq:ising-correlators}
\end{align}
Expanding the paired terms in~\eqref{eq:rotated-exact} cancels all odd powers of $O$, so the first correction is $4\binom{2\alpha}{2}E^{2\alpha-2}O^2$.
Since $E\to e_0$ and $O\sim\mathcal A_\sigma r^{-1/4}/2$, this gives~\eqref{eq:rotated-power}, while the next terms, $O^4$ and $E-e_0$, start at $\eta^{1/2}$ and $\eta$, respectively.

For the unrotated state,
\begin{equation}
 F_{\alpha,\eps} - 1 = \frac{|c_{xx}|^{2\alpha}+|c_{yy}|^{2\alpha}+|c_{zz}|^{2\alpha}-m_z^{4\alpha}}{(1+m_z^{2\alpha})^2} \,.
 \label{eq:unrotated-exact}
\end{equation}
The spin term has power $\eta^{\alpha/4}$ and the connected energy term has power $\eta$, both with positive coefficients, which proves~\eqref{eq:completion-test} without a numerical fit.

\section{Boundary data and the index two junction}
\label{app:boundary}

We review the boundary normalization and calculate the index two junction between the measured boundary and the transparent reference used in Section~\ref{sec:charged-amplitude}.

\subsection{Boundary normalization}
Consider $N$ folded fields with action $\kappa\int(\partial\phi)^2/(4\pi)$ and period $2\pi R$.
We impose Dirichlet conditions on $N-1$ relative fields and a Neumann condition on the collective field.
The overlap of this mixed boundary in the auxiliary circle theory and the fully Dirichlet overlap are~\cite{Hoshino:2025sre}
\begin{equation}
 g_{\text{mix},\text{circ}} = \sqrt{N/2}\,(\sqrt{2\kappa}R)^{1-N/2} \,,\qquad
 g_{\text{D},\text{circ}}=(\sqrt{2\kappa}R)^{-N/2} \,.
 \label{eq:circle-overlaps}
\end{equation}
Their ratio is $\sqrt{N/2}\sqrt{2\kappa}R$, and the oscillator BCC sector for this fully Dirichlet--mixed pair has one D--N twist of weight $1/16$.
This gluing does not follow from a generic Bell tensor, for which the relative field and dual field locking perturbations need not be minimized simultaneously.

For bosonized Ising we have $N=2\alpha$, $\kappa=2$ and $R=1$, which gives $\sqrt\alpha\,2^{1-\alpha}$ for the first overlap.
The $N$ independent reflections form $G=(\mathbb Z_2)^N$, and summing the fixed point mixed boundary over $G/\{\pm\id\}$ multiplies the overlap by $|G/\{\pm\id\}|/\sqrt{|G|}=2^{N/2-1}$, so that $g_{\text{mix},\text{orb}}=\sqrt\alpha$.
The regular Dirichlet boundary has $g_{\text{D},\text{orb}}=\sqrt{|G|}g_{\text{D},\text{circ}}=1$, and since the transparent reference also has unit vacuum overlap, the symmetric full system constant in~\eqref{eq:ising-g} is reproduced.
However, equal vacuum overlaps do not make the fully Dirichlet boundary and the transparent reference interchangeable in a junction calculation.
Other completions need their own projection and boundary map.
In particular, the transitions of the convolution based entropy in~\cite{Matsuda:2026} cannot be transferred to our setting without first showing that the two observables are equivalent.

\subsection{The index two orbifold junction}
\label{app:orbifold-junction}

We work in the elementary boundary family of~\cite{Hoshino:2025sre}, with displacement and Wilson line both set to zero, and let $\uparrow,\downarrow$ denote the upper and lower sides of the fold.
In the order $(\phi_1^\uparrow,\phi_2^\uparrow,\phi_1^\downarrow,\phi_2^\downarrow)$, the Neumann projectors for the transparent reference and the measured boundary are
\begin{equation}
 \mathsf P_0=\frac12\begin{pmatrix}1&0&1&0\\0&1&0&1\\1&0&1&0\\0&1&0&1\end{pmatrix} \,,
 \qquad \mathsf P_1=vv^T \,,\qquad v=\tfrac12(1,1,1,1)^T \,.
 \label{eq:boundary-projectors}
\end{equation}
We write $\mathsf R_a=2\mathsf P_a-\id$ for the corresponding reflection matrices.
Under the independent sign reflections $G=(\mathbb Z_2)^4$, the two boundaries have stabilizers $G_{\mathsf P_0}=\{\operatorname{diag}(s_1,s_2,s_1,s_2):s_1,s_2=\pm1\}$ and $G_{\mathsf P_1}=\{\id,-\id\}$.
An image of the measured boundary is represented by $s=\operatorname{diag}(s_1,s_2,s_3,s_4)$ with $s_i=\pm1$, and the four relative image sectors are labeled by $\delta_1=s_1s_3$ and $\delta_2=s_2s_4$.
The eigenvalues of $\mathsf R_0s\mathsf R_1s$ are $(1,1,1,-1)$ for $(\delta_1,\delta_2)=(+,+)$, $(1,-1,i,-i)$ when the signs differ, and $(1,-1,-1,-1)$ for $(-,-)$, with oscillator ground weights $1/16$, $5/32$ and $3/16$, respectively.
Since compact zero modes cannot lower these weights, the lowest junction lies in the matched $(+,+)$ sector, which has a unique zero momentum, zero stretching vacuum of weight $h_0=1/16$.

The two matched lifts, diagonal and anti-diagonal, form one orbit of $G_{\mathsf P_0}$, and their common stabilizer is generated by the total reflection $\mathsf J=-\id$.
The transparent reference has the positive fractional character, and we write $\xi_1=\pm1$ for the two resolutions of the measured boundary.
The open channel projection is then $(\id+\xi_1\mathsf J)/2$.
Since the lowest vacuum is $\mathsf J$-even, it survives once for $\xi_1=+1$ and is absent for $\xi_1=-1$, so the circle cover oscillator alone does not decide whether the lowest junction survives.

This sign is, however, constrained by the lattice projector.
Let $H_4$ be the four copy Hamiltonian, $|\Omega\rangle$ its vacuum of energy $E_0$, and $O=X_x^{(1)}X_x^{(3)}$ with $x$ in the interior of the interval.
Since $\mathscr D_A^2=\mathscr D_A=\mathscr D_A^\dagger$, we have
\begin{equation}
 \frac{\langle\Omega|O\ee^{-\tau(H_4-E_0)}\mathscr D_A \ee^{-\tau(H_4-E_0)}O|\Omega\rangle}{\langle\mathscr D_A\rangle} = \frac{\|\mathscr D_A\ee^{-\tau(H_4-E_0)}O|\Omega\rangle\|^2}{\langle\mathscr D_A\rangle} \geq 0 \,.
 \label{eq:mirror-projector-positivity}
\end{equation}
In the interior scaling regime $\eps\ll\tau\ll\operatorname{dist}(x,\partial A)$, the folded probe contains one spin in each bosonized pair and therefore lies in the fully twisted sector.
The normalized twisted ground components of the measured boundary and the transparent reference are $|t_1\rangle=(|0000\rangle+|1111\rangle)/\sqrt2$ and $|t_0\rangle=\frac12\sum_{a,b=0}^1|abab\rangle$, where $0,1$ denote the fixed points $0,\pi$.
If we write a unit single spin twist state as $|\mu\rangle=\mu_0|0\rangle+\mu_\pi|\pi\rangle$, its mirror overlap with $t_1$ is $(|\mu_0|^4+|\mu_\pi|^4)/\sqrt2>0$, while its overlap with $t_0$ is $(|\mu_0|^2+|\mu_\pi|^2)^2/2>0$.
The measured one-point function therefore has sign $\xi_1$, and the negative elementary resolution would violate~\eqref{eq:mirror-projector-positivity}.
Hence, the elementary boundary, if it is reached, has $\xi_1=+1$, although this argument does not exclude a direct sum of resolutions.

To calculate the charged coupling of this junction, we unfold~\eqref{eq:boundary-projectors}, after which only $\phi_-$ sees a Dirichlet slit.
The map
\begin{equation}
 \hat z(z)=\frac{2z-\ell+2\sqrt{z(z-\ell)}}{\ell} \,,\qquad
 \frac{\hat z'(r)}{\hat z(r)^2-1}=\frac{\ell}{4r(r-\ell)} \,,\quad (r>\ell) \,,
 \label{eq:slit-map-v2}
\end{equation}
takes the slit plane to the exterior unit disc, with $\hat z(z)\sim4z/\ell$ at infinity.
A unit vertex of dimension $\Delta$ has the Dirichlet one-point function $(2\operatorname{Im}z)^{-\Delta}$ in the upper half-plane, whose transform is $[|\hat z'|/(|\hat z|^2-1)]^\Delta$.
For the relative cosine in~\eqref{eq:charged-bosonization} we have $\Delta=1/2$, which gives~\eqref{eq:slit-response}, while the collective cosine has zero expectation.
Local endpoint factors and the conformal anomaly cancel between the numerator and the denominator.

The charged field is orbifold invariant and untwisted, and it preserves the covering image with dimensionless response $1/2$ on both matched lifts.
On their two-dimensional space, the remaining image projection and response are
\begin{equation}
 \Pi = \frac12\begin{pmatrix}1&1\\1&1\end{pmatrix} \,,\qquad
 \mathsf K_{\mathcal S} = \tfrac12\id_2 \,,\qquad
 b_\text{orb} = \frac{\Tr(\Pi\mathsf K_{\mathcal S})}{\Tr\Pi} = \frac12 \,.
 \label{eq:orbifold-coupling}
\end{equation}
The common stabilizer projection has already retained the even vacuum, so the image projection does not double the charged contribution.
Squaring this coupling gives $1/4$.

A remaining obstruction to identifying the microscopic endpoint is that orbifold averaging does not commute with exponentiating a boundary action.
If $\mathscr P_\text{refl}$ averages the independent sign reflections, then at a fixed field regulator
\begin{equation}
 \mathscr P_\text{refl}\ee^{\lambda\cos(\phi_1-\phi_2)} = \ee^{\lambda\cos\phi_1\cos\phi_2}\cosh(\lambda\sin\phi_1\sin\phi_2)
 \neq \ee^{\lambda\cos\phi_1\cos\phi_2}
 = \ee^{\lambda\mathscr P_\text{refl}\cos(\phi_1-\phi_2)} \,,
 \label{eq:orbifold-average-caveat}
\end{equation}
in general.
Agreement of reflected linear operators therefore does not imply agreement of their boundary weights.

\subsection{Zero modes and topological lines}
\label{app:zero-modes-lines}

\paragraph{Zero modes:}
To see the role of zero modes, let us consider four D--N twists of weight $1/16$ in an auxiliary noncompact boson, whose normalized oscillator correlator is~\cite{Mattiello:2018twist,Erler:2014bcc}
\begin{equation}
 \mathfrak F_\text{DN}^\text{osc}(\eta) = (1-\eta)^{-1/8}\left[\frac2\pi K(\sqrt\eta)\right]^{-1/2}
 = 1 + \frac{\eta^2}{128}+O(\eta^3) \,,
 \label{eq:DN}
\end{equation}
where $K$ is the complete elliptic integral of the first kind in the modulus convention.
The order $\eta$ terms cancel, which leaves the stress coefficient $2h^2/c=1/128$ for $h=1/16$ and auxiliary central charge $c=1$.
With normalization $\kappa=1$, compactification multiplies the result by the Jacobi theta function $\vartheta_3(0,q^{R^2})$, where $q=\exp[-\pi K(\sqrt{1-\eta})/K(\sqrt\eta)]$.
At $R=1$, however, the theta identity reduces the product to $(1-\eta)^{-1/8}$ and so restores an order $\eta$ term.
The oscillator calculation alone therefore cannot determine which neutral terms survive the contraction at the Ising orbifold junction.

\paragraph{Topological lines:}
A Bell boundary is not automatically a topological line.
Consider a rational, algebra preserving boundary with Ishibashi coefficients $\Psi_{\Gamma j}$.
Let $S_{ij}$ be the modular matrix with $0$ labeling the vacuum representation, and suppose that a separately identified Verlinde line $\mathcal D_x$ acts with eigenvalue $S_{xj}/S_{0j}$.
Performing a modular transformation on the closed channel characters of the defect threaded annulus then gives the open channel coefficients
\begin{equation}
 n_{\Gamma,\mathcal D_x\Gamma}^{\ i} = \sum_j|\Psi_{\Gamma j}|^2\frac{S_{xj}}{S_{0j}}S_{ji} \,.
 \label{eq:topological-annulus}
\end{equation}
For a Cardy boundary, substituting $\Psi_{aj}=S_{aj}/\sqrt{S_{0j}}$ and applying the Verlinde formula gives the usual fusion action~\cite{Cardy:1989ir}, whereas for a Bell boundary the coefficients and the line must first be identified~\cite{Hoshino:2025fusion}.

\clearpage
\section{Notation}
\label{app:notation}

Table~\ref{tab:notation} collects the main symbols and indicates where they are defined.

\begin{table}[!htbp]
\centering\small\renewcommand{\arraystretch}{1.0}
\begin{tabular}{@{}>{\raggedright\arraybackslash}p{0.20\textwidth}>{\raggedright\arraybackslash}p{0.58\textwidth}>{\raggedright\arraybackslash}p{0.16\textwidth}@{}}
\hline
Symbol & Meaning & Defined in\\
\hline
\multicolumn{3}{@{}l}{\emph{Moments and two region ratios}}\\
$W_{\bm u}$, $\chi_A(\bm u)$ & Weyl operator and its expectation $\Tr(\rho_AW_{\bm u})$ &~\eqref{eq:weyl},~\eqref{eq:moment}\\
$\Xi_A(\bm u)$, $Z_\alpha(A)$ & participation weight $|\chi_A|^2/D_A$ and moment $\sum_{\bm u}\Xi_A^\alpha$ &~\eqref{eq:moment}\\
$P_A$, $S_2(A)$ & purity $\Tr\rho_A^2$ and R\'enyi-$2$ entropy $-\log P_A$ &~\eqref{eq:intro-baseline}, Section~\ref{sec:observable}\\
$\widehat\M_\alpha(A)$ & stabilizer R\'enyi functional of a reduced state &~\eqref{eq:mhat}\\
$\mathcal A_\alpha(A)$ & residual beyond the entanglement baseline &~\eqref{eq:residual}\\
$F_{\alpha,\eps}$, $\I_\alpha^\text{stab}$, $I_2$ & two region ratio, normalized logarithm $\log F_{\alpha,\eps}/(\alpha-1)$, and R\'enyi-$2$ mutual information &~\eqref{eq:ratio}\\
$\eta$ & cross-ratio, $\eta\simeq\ell_A\ell_B/r^2$ for distant intervals &~\eqref{eq:eta}\\
\multicolumn{3}{@{}l}{\emph{Charges and replicas}}\\
$q(\bm u)$ & charge of a Weyl label under $\prod_jX_j$ &~\eqref{eq:charge}\\
$Z_\alpha^{\text{ch},1}$ & charged sum over the fundamental charges $q=\pm1$ & Section~\ref{sec:selection}\\
$h_\text{ch}$, $\Delta^\text{ch}_{\min}$ & charged exponent and lightest charged dimension &~\eqref{eq:hcharged}\\
$K_\alpha$ & replica subgroup preserved by the Bell tensor &~\eqref{eq:replica-symmetry}\\
\multicolumn{3}{@{}l}{\emph{Boundary description}}\\
$\Gamma_1$, $\Gamma_0$, $\Gamma_2$ & measured boundary, transparent reference, state preparation boundary & Section~\ref{sec:boundary}\\
$\psi_{ab}$, $h_\alpha$ & BCC field from $\Gamma_a$ to $\Gamma_b$ and leading endpoint weight & Section~\ref{sec:boundary}\\
$h_\star$ & leading exchanged weight in the reference channel &~\eqref{eq:blocks}\\
$A_1$ & coefficient of the linear neutral power $\eta^{\Delta_\varepsilon}$ &~\eqref{eq:neutral-overlap}\\
$\mathcal C_\alpha$ & coefficient of the charged power $\eta^{h_\text{ch}}$ &~\eqref{eq:uniform},~\eqref{eq:charged-coefficient}\\
(A1)--(A4) & sufficient assumptions for the candidate continuum exponent & Section~\ref{sec:conditions}\\
(A5) & alternative kink matching hypothesis, incompatible with (A2) on the locked branch & Section~\ref{sec:kink}\\
\multicolumn{3}{@{}l}{\emph{Exact Ising sums}}\\
$\mathcal A_\sigma$, $m_z$ & amplitude in $\langle X_0X_r\rangle\sim\mathcal A_\sigma r^{-1/4}$, and $\langle Z\rangle=2/\pi$ &~\eqref{eq:rotated-power}\\
$T_\alpha(\ell)$ & $\sum_{\bm u}|\chi_A(\bm u)|^{2\alpha}$ for an $\ell$-site interval &~\eqref{eq:neutral-overlap}\\
$C_{\bm u}$, $\Sigma_\alpha$, $U_\alpha$ & charged overlap, charged sum, and $U_\alpha=\Sigma_\alpha/T_\alpha$ &~\eqref{eq:U}\\
$\mathscr Z_N^{(\alpha)}[w]$, $w_0$, $w_1$ & ensemble on $\Lambda_\ell$, with neutral and charged weights &~\eqref{eq:ensemble},~\eqref{eq:weights}\\
$\mathsf M^{(N)}[w]$ & skew moment matrix, with $\Pf\mathsf M^{(N)}=\mathscr Z_N^{(2)}$ &~\eqref{eq:skew-moments}\\
$\mathscr D_A$ & index two lattice projector $\prod_{x\in A}Q_{2,x}$ &~\eqref{eq:Q2-projector}\\
$b_\ell$ & far field charged coupling, with $\mathcal C_2=b_\ell^2$ &~\eqref{eq:charged-coupling}\\
\hline
\end{tabular}
\caption{\textsf{
Main notation, grouped by topic.}}
\label{tab:notation}
\end{table}
\FloatBarrier

\begingroup
\small\renewcommand{\baselinestretch}{1.0}\selectfont
\bibliographystyle{JHEP}
\bibliography{refs}

@article{White:2020zoz,
    author = "White, Christopher David and Cao, ChunJun and Swingle, Brian",
    title = "{Conformal field theories are magical}",
    eprint = "2007.01303",
    archivePrefix = "arXiv",
    primaryClass = "quant-ph",
    doi = "10.1103/PhysRevB.103.075145",
    journal = "Phys. Rev. B",
    volume = "103",
    number = "7",
    pages = "075145",
    year = "2021"
}

@article{Cao:2023fcm,
    author = "Cao, ChunJun",
    title = "{Non-trivial Area Operators Require Non-local Magic}",
    eprint = "2306.14996",
    archivePrefix = "arXiv",
    primaryClass = "hep-th",
    doi = "10.1007/JHEP11(2024)105",
    journal = "JHEP",
    volume = "11",
    pages = "105",
    year = "2024"
}

@article{Cao:2024gdi,
    author = "Cao, ChunJun and Cheng, Gong and Hamma, Alioscia and Leone, Lorenzo and Munizzi, William and Oliviero, Salvatore F. E.",
    title = "{Gravitational back-reaction is magical}",
    eprint = "2403.07056",
    archivePrefix = "arXiv",
    primaryClass = "hep-th",
    doi = "10.1103/z3vr-w5c5",
    journal = "PRX Quantum",
    volume = "6",
    number = "4",
    pages = "040375",
    year = "2025"
}

@article{Hoshino:2025sre,
    author = "Hoshino, Masahiro and Oshikawa, Masaki and Ashida, Yuto",
    title = "{Stabilizer R\'enyi Entropy and Conformal Field Theory}",
    eprint = "2503.13599",
    archivePrefix = "arXiv",
    primaryClass = "quant-ph",
    doi = "10.1103/ylsz-dm3y",
    journal = "Phys. Rev. X",
    volume = "16",
    number = "1",
    pages = "011037",
    year = "2026"
}

@article{Hoshino:2025fusion,
    author = "Hoshino, Masahiro and Ashida, Yuto",
    title = "{Stabilizer R\'enyi Entropy Encodes Fusion Rules of Topological Defects and Boundaries}",
    eprint = "2507.10656",
    archivePrefix = "arXiv",
    primaryClass = "quant-ph",
    doi = "10.1103/1tyr-rlbb",
    journal = "Phys. Rev. Lett.",
    volume = "136",
    number = "8",
    pages = "080402",
    year = "2026"
}

@article{Malvimat:2026syk,
    author = "Malvimat, Vinay and Sarkis, Matthieu and Suk, Yena and Yoon, Junggi",
    title = "{Multipartite Non-local Magic and SYK Model}",
    eprint = "2601.03076",
    archivePrefix = "arXiv",
    primaryClass = "hep-th",
    year = "2026"
}

@article{Cao:2026codes,
    author = "Cao, ChunJun and Cheng, Gong and Karthikeyan, Krishnanand and Li, Cathy and Preskill, John",
    title = "{State-dependent geometries from magic-enriched quantum codes}",
    eprint = "2603.13475",
    archivePrefix = "arXiv",
    primaryClass = "hep-th",
    year = "2026"
}

@article{Leone:2021sre,
    author = "Leone, Lorenzo and Oliviero, Salvatore F. E. and Hamma, Alioscia",
    title = "{Stabilizer R\'enyi Entropy}",
    eprint = "2106.12587",
    archivePrefix = "arXiv",
    primaryClass = "quant-ph",
    doi = "10.1103/PhysRevLett.128.050402",
    journal = "Phys. Rev. Lett.",
    volume = "128",
    number = "5",
    pages = "050402",
    year = "2022"
}

@article{Veitch:2014rwa,
    author = "Veitch, Victor and Mousavian, S. A. Hamed and Gottesman, Daniel and Emerson, Joseph",
    title = "{The resource theory of stabilizer quantum computation}",
    eprint = "1307.7171",
    archivePrefix = "arXiv",
    primaryClass = "quant-ph",
    doi = "10.1088/1367-2630/16/1/013009",
    journal = "New J. Phys.",
    volume = "16",
    pages = "013009",
    year = "2014"
}

@article{Howard:2014qwa,
    author = "Howard, Mark and Wallman, Joel and Veitch, Victor and Emerson, Joseph",
    title = "{Contextuality supplies the magic for quantum computation}",
    eprint = "1401.4174",
    archivePrefix = "arXiv",
    primaryClass = "quant-ph",
    doi = "10.1038/nature13460",
    journal = "Nature",
    volume = "510",
    pages = "351--355",
    year = "2014"
}

@article{Gross:2006,
    author = "Gross, David",
    title = "{Hudson's theorem for finite-dimensional quantum systems}",
    eprint = "quant-ph/0602001",
    archivePrefix = "arXiv",
    doi = "10.1063/1.2393152",
    journal = "J. Math. Phys.",
    volume = "47",
    pages = "122107",
    year = "2006"
}

@article{Fateev:1985para,
    author = "Zamolodchikov, Alexander B. and Fateev, Vladimir A.",
    title = "{Nonlocal (parafermion) currents in two-dimensional conformal quantum field theory and self-dual critical points in $Z_N$-symmetric statistical systems}",
    journal = "Sov. Phys. JETP",
    volume = "62",
    number = "2",
    pages = "215--225",
    year = "1985"
}

@article{Hostens:2005,
    author = "Hostens, Erik and Dehaene, Jeroen and De Moor, Bart",
    title = "{Stabilizer states and Clifford operations for systems of arbitrary dimensions, and modular arithmetic}",
    eprint = "quant-ph/0408190",
    archivePrefix = "arXiv",
    doi = "10.1103/PhysRevA.71.042315",
    journal = "Phys. Rev. A",
    volume = "71",
    pages = "042315",
    year = "2005"
}

@article{Recknagel:2002perm,
    author = "Recknagel, Andreas",
    title = "{Permutation branes}",
    eprint = "hep-th/0208119",
    archivePrefix = "arXiv",
    doi = "10.1088/1126-6708/2003/04/041",
    journal = "JHEP",
    volume = "04",
    pages = "041",
    year = "2003"
}

@article{Cardy:1989ir,
    author = "Cardy, John L.",
    title = "{Boundary Conditions, Fusion Rules and the Verlinde Formula}",
    doi = "10.1016/0550-3213(89)90521-X",
    journal = "Nucl. Phys. B",
    volume = "324",
    pages = "581--596",
    year = "1989"
}

@article{Billo:2016defects,
    author = "Bill\`o, Marco and Gon\c{c}alves, Vasco and Lauria, Edoardo and Meineri, Marco",
    title = "{Defects in conformal field theory}",
    eprint = "1601.02883",
    archivePrefix = "arXiv",
    primaryClass = "hep-th",
    doi = "10.1007/JHEP04(2016)091",
    journal = "JHEP",
    volume = "04",
    pages = "091",
    year = "2016"
}

@article{Affleck:1991tk,
    author = "Affleck, Ian and Ludwig, Andreas W. W.",
    title = "{Universal noninteger ground-state degeneracy in critical quantum systems}",
    doi = "10.1103/PhysRevLett.67.161",
    journal = "Phys. Rev. Lett.",
    volume = "67",
    pages = "161--164",
    year = "1991"
}

@article{Calabrese:2004eu,
    author = "Calabrese, Pasquale and Cardy, John L.",
    title = "{Entanglement entropy and quantum field theory}",
    eprint = "hep-th/0405152",
    archivePrefix = "arXiv",
    doi = "10.1088/1742-5468/2004/06/P06002",
    journal = "J. Stat. Mech.",
    volume = "0406",
    pages = "P06002",
    year = "2004"
}

@article{Calabrese:2009qy,
    author = "Calabrese, Pasquale and Cardy, John L.",
    title = "{Entanglement entropy and conformal field theory}",
    eprint = "0905.4013",
    archivePrefix = "arXiv",
    primaryClass = "cond-mat.stat-mech",
    doi = "10.1088/1751-8113/42/50/504005",
    journal = "J. Phys. A",
    volume = "42",
    pages = "504005",
    year = "2009"
}

@article{CardyCalabrese:2010corr,
    author = "Cardy, John and Calabrese, Pasquale",
    title = "{Unusual corrections to scaling in entanglement entropy}",
    eprint = "1002.4353",
    archivePrefix = "arXiv",
    primaryClass = "cond-mat.stat-mech",
    doi = "10.1088/1742-5468/2010/04/P04023",
    journal = "J. Stat. Mech.",
    volume = "2010",
    number = "04",
    pages = "P04023",
    year = "2010"
}

@article{Ryu:2006bv,
    author = "Ryu, Shinsei and Takayanagi, Tadashi",
    title = "{Holographic derivation of entanglement entropy from AdS/CFT}",
    eprint = "hep-th/0603001",
    archivePrefix = "arXiv",
    doi = "10.1103/PhysRevLett.96.181602",
    journal = "Phys. Rev. Lett.",
    volume = "96",
    pages = "181602",
    year = "2006"
}

@article{Hubeny:2007xt,
    author = "Hubeny, Veronika E. and Rangamani, Mukund and Takayanagi, Tadashi",
    title = "{A Covariant holographic entanglement entropy proposal}",
    eprint = "0705.0016",
    archivePrefix = "arXiv",
    primaryClass = "hep-th",
    doi = "10.1088/1126-6708/2007/07/062",
    journal = "JHEP",
    volume = "07",
    pages = "062",
    year = "2007"
}

@article{Faulkner:2013ana,
    author = "Faulkner, Thomas and Lewkowycz, Aitor and Maldacena, Juan",
    title = "{Quantum corrections to holographic entanglement entropy}",
    eprint = "1307.2892",
    archivePrefix = "arXiv",
    primaryClass = "hep-th",
    doi = "10.1007/JHEP11(2013)074",
    journal = "JHEP",
    volume = "11",
    pages = "074",
    year = "2013"
}

@article{Engelhardt:2014gca,
    author = "Engelhardt, Netta and Wall, Aron C.",
    title = "{Quantum Extremal Surfaces: Holographic Entanglement Entropy beyond the Classical Regime}",
    eprint = "1408.3203",
    archivePrefix = "arXiv",
    primaryClass = "hep-th",
    doi = "10.1007/JHEP01(2015)073",
    journal = "JHEP",
    volume = "01",
    pages = "073",
    year = "2015"
}

@article{Harlow:2016vwg,
    author = "Harlow, Daniel",
    title = "{The Ryu-Takayanagi Formula from Quantum Error Correction}",
    eprint = "1607.03901",
    archivePrefix = "arXiv",
    primaryClass = "hep-th",
    doi = "10.1007/s00220-017-2904-z",
    journal = "Commun. Math. Phys.",
    volume = "354",
    number = "3",
    pages = "865--912",
    year = "2017"
}

@article{Stephan:2011rse,
    author = "St\'ephan, Jean-Marie and Misguich, Gr\'egoire and Pasquier, Vincent",
    title = "{Phase transition in the R\'enyi-Shannon entropy of Luttinger liquids}",
    journal = "Phys. Rev. B",
    volume = "84",
    pages = "195128",
    year = "2011",
    doi = "10.1103/PhysRevB.84.195128",
    eprint = "1104.2544",
    archivePrefix = "arXiv"
}

@article{LeoneBittel:2024,
    author = "Leone, Lorenzo and Bittel, Lennart",
    title  = "{Stabilizer entropies are monotones for magic-state resource theory}",
    journal = "Phys. Rev. A",
    volume = "110",
    pages  = "L040403",
    year   = "2024",
    doi = "10.1103/PhysRevA.110.L040403",
    eprint = "2404.11652",
    archivePrefix = "arXiv"
}

@article{Affleck:1998potts,
    author = "Affleck, Ian and Oshikawa, Masaki and Saleur, Hubert",
    title  = "{Boundary critical phenomena in the three-state Potts model}",
    journal = "J. Phys. A",
    volume = "31",
    pages  = "5827--5842",
    year   = "1998",
    doi = "10.1088/0305-4470/31/28/003",
    eprint = "cond-mat/9804117",
    archivePrefix = "arXiv"
}

@article{Mong:2014,
    author = "Mong, Roger S. K. and Clarke, David J. and Alicea, Jason and Lindner, Netanel H. and Fendley, Paul",
    title  = "{Parafermionic conformal field theory on the lattice}",
    journal = "J. Phys. A",
    volume = "47",
    pages  = "452001",
    year   = "2014",
    doi = "10.1088/1751-8113/47/45/452001",
    eprint = "1406.0846",
    archivePrefix = "arXiv"
}

@article{Benedetti:2026,
    author = "Benedetti, Valentin and Dabholkar, Atish and Dalmonte, Marcello",
    title  = "{Universality of Magic in Local Quantum Field Theory}",
    year   = "2026",
    eprint = "2607.16403",
    archivePrefix = "arXiv"
}

@article{Matsuda:2026,
    author = "Matsuda, Ryota and Hoshino, Masahiro and Ashida, Yuto",
    title  = "{Quantum Computational Resources and Conformal Field Theory: Unifying Spins, Bosons, and Fermions}",
    year   = "2026",
    eprint = "2607.05343",
    archivePrefix = "arXiv"
}

@article{Rajabpour:2025equiv,
    author = "Ramirez Trino, E. A. and Rajabpour, M. A.",
    title  = "{Equivalence of Stabilizer and Shannon R\'enyi Entropies: Exact Results for Quantum Critical Chains}",
    journal = "Phys. Rev. Lett.",
    volume = "137",
    pages  = "090403",
    doi    = "10.1103/2frt-tdg9",
    year   = "2026",
    eprint = "2509.10700",
    archivePrefix = "arXiv"
}

@article{Khasseh:2026fugacity,
    author = "Khasseh, Reyhaneh and Rajabpour, M. A.",
    title  = "{Fugacity-Resolved Stabilizer Entropy in Critical Quantum Chains: Discrete Selberg Sums and Exactly Solvable R\'enyi Indices}",
    year   = "2026",
    eprint = "2608.06995",
    archivePrefix = "arXiv"
}

@article{Stephan:2014mi,
    doi = "10.1103/PhysRevB.90.045424",
    author = "St\'ephan, Jean-Marie",
    title  = "{Shannon and R\'enyi mutual information in quantum critical spin chains}",
    journal = "Phys. Rev. B",
    volume = "90",
    pages  = "045424",
    year   = "2014",
    eprint = "1403.6157",
    archivePrefix = "arXiv"
}

@article{Alcaraz:2014mi,
    doi = "10.1103/PhysRevB.90.075132",
    author = "Alcaraz, F. C. and Rajabpour, M. A.",
    title  = "{Universal behavior of the Shannon and R\'enyi mutual information of quantum critical chains}",
    journal = "Phys. Rev. B",
    volume = "90",
    pages  = "075132",
    year   = "2014",
    eprint = "1405.1074",
    archivePrefix = "arXiv"
}

@article{Dong:2016fnf,
    author = "Dong, Xi",
    title  = "{The Gravity Dual of R\'enyi Entropy}",
    journal = "Nature Commun.",
    volume = "7",
    pages  = "12472",
    year   = "2016",
    doi = "10.1038/ncomms12472",
    eprint = "1601.06788",
    archivePrefix = "arXiv",
    primaryClass = "hep-th"
}

@article{Hayden:2011ag,
    author = "Hayden, Patrick and Headrick, Matthew and Maloney, Alexander",
    title  = "{Holographic mutual information is monogamous}",
    journal = "Phys. Rev. D",
    volume = "87",
    pages  = "046003",
    year   = "2013",
    doi = "10.1103/PhysRevD.87.046003",
    eprint = "1107.2940",
    archivePrefix = "arXiv",
    primaryClass = "hep-th"
}

@article{Fattal:2004stabilizer,
    author = "Fattal, David and Cubitt, Toby S. and Yamamoto, Yoshihisa and Bravyi, Sergey and Chuang, Isaac L.",
    title = "{Entanglement in the stabilizer formalism}",
    eprint = "quant-ph/0406168",
    archivePrefix = "arXiv",
    primaryClass = "quant-ph",
    year = "2004"
}

@article{Haug:2023monotones,
    doi = "10.22331/q-2023-08-28-1092",
    author = "Haug, Tobias and Piroli, Lorenzo",
    title = "{Stabilizer entropies and nonstabilizerness monotones}",
    eprint = "2303.10152",
    archivePrefix = "arXiv",
    primaryClass = "quant-ph",
    journal = "Quantum",
    volume = "7",
    pages = "1092",
    year = "2023"
}

@article{Tirrito:2024flatness,
    doi = "10.1103/PhysRevA.109.L040401",
    author = "Tirrito, Emanuele and Tarabunga, Poetri Sonya and Lami, Guglielmo and Chanda, Titas and Leone, Lorenzo and Oliviero, Salvatore F. E. and Dalmonte, Marcello and Collura, Mario and Hamma, Alioscia",
    title = "{Quantifying non-stabilizerness through entanglement spectrum flatness}",
    eprint = "2304.01175",
    archivePrefix = "arXiv",
    primaryClass = "quant-ph",
    journal = "Phys. Rev. A",
    volume = "109",
    pages = "L040401",
    year = "2024"
}

@article{Tarabunga:2023longrange,
    doi = "10.1103/PRXQuantum.4.040317",
    author = "Tarabunga, Poetri Sonya and Tirrito, Emanuele and Chanda, Titas and Dalmonte, Marcello",
    title = "{Many-body magic via Pauli-Markov chains---from criticality to gauge theories}",
    eprint = "2305.18541",
    archivePrefix = "arXiv",
    primaryClass = "quant-ph",
    journal = "PRX Quantum",
    volume = "4",
    pages = "040317",
    year = "2023"
}

@article{Korbany:2025longrange,
    doi = "10.1103/1hlj-h6t9",
    author = "Korbany, David Aram and Gullans, Michael J. and Piroli, Lorenzo",
    title = "{Long-range nonstabilizerness and phases of matter}",
    eprint = "2502.19504",
    archivePrefix = "arXiv",
    primaryClass = "quant-ph",
    journal = "Phys. Rev. Lett.",
    volume = "135",
    pages = "160404",
    year = "2025"
}

@article{Haug:2023bell,
    doi = "10.1103/PRXQuantum.4.010301",
    author = "Haug, Tobias and Kim, M. S.",
    title = "{Scalable measures of magic resource for quantum computers}",
    eprint = "2204.10061",
    archivePrefix = "arXiv",
    primaryClass = "quant-ph",
    journal = "PRX Quantum",
    volume = "4",
    pages = "010301",
    year = "2023"
}

@article{Haug:2023mps,
    doi = "10.1103/PhysRevB.107.035148",
    author = "Haug, Tobias and Piroli, Lorenzo",
    title = "{Quantifying nonstabilizerness of matrix product states}",
    eprint = "2207.13076",
    archivePrefix = "arXiv",
    primaryClass = "quant-ph",
    journal = "Phys. Rev. B",
    volume = "107",
    pages = "035148",
    year = "2023"
}

@article{Xavier:2011bc,
    doi = "10.1103/PhysRevB.84.094410",
    author = "Xavier, J. C. and Alcaraz, F. C.",
    title = "{Precise determination of quantum critical points by the violation of the entropic area law}",
    eprint = "1106.4527",
    archivePrefix = "arXiv",
    primaryClass = "cond-mat.stat-mech",
    journal = "Phys. Rev. B",
    volume = "84",
    pages = "094410",
    year = "2011"
}

@article{Chang:2019tdl,
    author = "Chang, Chi-Ming and Lin, Ying-Hsuan and Shao, Shu-Heng and Wang, Yifan and Yin, Xi",
    title = "{Topological Defect Lines and Renormalization Group Flows in Two Dimensions}",
    eprint = "1802.04445",
    archivePrefix = "arXiv",
    primaryClass = "hep-th",
    doi = "10.1007/JHEP01(2019)026",
    journal = "JHEP",
    volume = "01",
    pages = "026",
    year = "2019"
}

@article{Seiberg:2024kw,
    author = "Seiberg, Nathan and Shao, Shu-Heng",
    title = "{Majorana chain and Ising model---(non-invertible) translations, anomalies, and emanant symmetries}",
    eprint = "2307.02534",
    archivePrefix = "arXiv",
    primaryClass = "cond-mat.str-el",
    doi = "10.21468/SciPostPhys.16.3.064",
    journal = "SciPost Phys.",
    volume = "16",
    number = "3",
    pages = "064",
    year = "2024"
}

@article{Mattiello:2018twist,
    author = "Mattiello, Luca and Sachs, Ivo",
    title = "{{$\mathbb Z_2$} boundary twist fields and the moduli space of {D}-branes}",
    eprint = "1803.07500",
    archivePrefix = "arXiv",
    primaryClass = "hep-th",
    doi = "10.1007/JHEP07(2018)099",
    journal = "JHEP",
    volume = "07",
    pages = "099",
    year = "2018"
}

@article{Erler:2014bcc,
    author = "Erler, Theodore and Maccaferri, Carlo",
    title = "{String Field Theory Solution for Any Open String Background}",
    eprint = "1406.3021",
    archivePrefix = "arXiv",
    primaryClass = "hep-th",
    doi = "10.1007/JHEP10(2014)029",
    journal = "JHEP",
    volume = "10",
    pages = "029",
    year = "2014"
}

@article{Chu:2026euw,
    author = "Chu, Jinwei and Kharel, Savan",
    title = "{Stabilizer R{\'e}nyi Entropy Without a Phase Reference}",
    eprint = "2610.18955",
    archivePrefix = "arXiv",
    primaryClass = "quant-ph",
    month = "10",
    year = "2026"
}

@article{Shaposhnik:2026inf,
    author = "Shaposhnik, Leo and Leone, Lorenzo and Bittel, Lennart and Saraidaris, Dimitris and Jahn, Alexander and Eisert, Jens",
    title = "{Infinite magic and the emergence of the continuum}",
    eprint = "2610.20790",
    archivePrefix = "arXiv",
    primaryClass = "quant-ph",
    month = "10",
    year = "2026"
}
\endgroup

\end{document}